\documentclass[fleqn,prepint]{elsarticle}
\usepackage{subfloat}
\usepackage{float}
\usepackage[export]{adjustbox}
\usepackage[parfill]{parskip}
\usepackage{pdfpages}
\usepackage{amsbsy}
\usepackage{amsfonts}
\usepackage{amssymb}
\usepackage{amsmath}
\usepackage{multirow}
\usepackage{caption}
\usepackage{subcaption}
\usepackage{graphicx}
\usepackage{rotating}
\usepackage{breqn}
\usepackage[toc,page,header]{appendix}
\usepackage[margin=1.0in]{geometry}
\usepackage[normalem]{ulem}
\usepackage{tabulary}
\usepackage{hyperref}
\usepackage{booktabs}
\hypersetup{
colorlinks = true,
allbordercolors = {white},
allcolors = {blue},
}
\journal{}

\allowdisplaybreaks

\newcommand{\stkout}[1]{\ifmmode\text{\sout{\ensuremath{#1}}}\else\sout{#1}\fi}

\newcommand{\mathleft}{\@fleqntrue\@mathmargin0pt}
\newcommand{\mathcenter}{\@fleqnfalse}
\begin{document}
\begin{frontmatter}
\title{Quantifying Resolutions for DNS and LES for Lax-Wendroff Method: Application to Uniform/Non-Uniform Compact Schemes}

\author[nal_addr,iitk_addr]{V.~K.~Suman}
\ead{vksuman@iitk.ac.in}

\author[iitk_addr]{P.~Sundaram}
\ead{prasanna@iitk.ac.in}

\author[fr_addr]{Soumyo~Sengupta}
\ead{soumyoss08@gmail.com}

\author[iitdh_addr]{Aditi~Sengupta}
\ead{aditi@iitism.ac.in}

\author[iitdh_addr]{Tapan~K.~Sengupta}
\ead{tksengupta@iitism.ac.in}

\address[nal_addr]{Computational \& Theoretical Fluid Dynamics Division, CSIR-NAL, Bengaluru-560017, India}
\address[iitk_addr]{High Performance Computing Laboratory, Dept. Aerospace Engineering, Indian Institute of Technology Kanpur, Kanpur-208016, India}
\address[fr_addr]{CERFACS, 42 Avenue G. Coriolis, 31057, Toulouse Cedex 1, France}
\address[iitdh_addr]{Dept. of Mechanical Engineering, Indian Institute of Technology (ISM) Dhanbad, Dhanbad-826004, India}

\begin{abstract}
The global spectral analysis (GSA) of numerical methods ensures that the dispersion relation preserving (DRP) property is calibrated in addition to ensuring numerical stability, as advocated in the von Neumann analysis. The DRP nature plays a major role where spatio-temporal dependence in the governing equation and boundary conditions has to be retained, such as in direct numerical simulations (DNS) and large eddy simulations (LES) of fluid flow transition. Using the concept of GSA, methods based on the Lax-Wendroff approach for temporal integration are calibrated using a high accuracy, sixth order non-uniform compact scheme, developed in \lq \lq Hybrid sixth order spatial discretization scheme for non-uniform Cartesian grids - Sharma {\it et al.} {\it Comput. Fluids}, {\bf 157}, 208-231 (2017)." The model equation used for this analysis is the one-dimensional (1D) convection-diffusion equation (CDE) which provides a unique state for the Lax-Wendroff method, results of which will have direct consequences for the solution of Navier-Stokes equations. Furthermore, the specific choice of the governing equation enables a direct assessment of the performance of numerical methods for solving fluid flows due to its one-to-one correspondence with the Navier-Stokes equation as established in ``Effects of numerical anti-diffusion in closed unsteady flows governed by two-dimensional Navier-Stokes equation- Suman {\it et al.} {\it Comput. Fluids}, {\bf 201}, 104479 (2020)''. The limiting case of the non-uniform compact scheme, which is a uniform grid, is considered. This is also investigated using GSA, and potential differences for the non-uniformity of grid are compared. Finally, further use of this newly developed Lax-Wendroff method for the non-uniformity of grid is quantified for its application in DNS and LES. 
\end{abstract}


\end{frontmatter}

\section{Introduction}
High performance computing (HPC) is becoming important with faster and more resolved solutions reported that help in understanding many previously unsolved problems. For example, three-dimensional (3D) transitional and turbulent flow over a shear layer has been solved from the receptivity stage to the fully developed turbulent state \cite{B64,B113,C20,C14} and other references contained therein. This has been a benchmark problem for some time now in DNS and LES of fluid flow transition and turbulence. The development of methods for DNS and LES has been critically assessed in \cite{C14}.

In \cite{B113,C20}, the authors use a domain decomposition based parallel algorithm to establish that the presence of spatio-temporal wave-fronts are responsible for creating transitional and turbulent flows. The 3D field has been computed using a compact scheme for the DNS performed on a staggered grid. The use of compact schemes is desirable for DNS and LES as these yield high accuracy while using fewer points, as compared to non-compact scheme based methods for non-periodic problems. The implicit nature of compact schemes enables a wider scale of resolution as compared to explicit schemes. Accurate numerical simulations of the Navier-Stokes
equation require resolution of spatial and temporal scales of flow as viewed in the spectral plane, rather than merely inspecting the order of the schemes by Taylor series expansion \cite{HACM153, HACM, HACM340, Gaitonde}. The compact schemes provide near-spectral resolution for the first spatial derivative, which is noted in the literature, obtained by full domain analysis using global spectral analysis (GSA) \cite{B18,Gaitonde}.

Compact schemes which function as finite difference methods are preferred for the higher accuracy they impart to the solutions. These schemes use implicit stencils to compute derivatives with \lq near-spectral \rq resolution while using relatively coarser grids. Most compact schemes for partial differential equations (PDEs) originate from the Pad\'e discretization schemes for ordinary differential equations (ODEs). There are many compact schemes \cite{HACM153,HACM,HACM4,HACM3,Gaitonde,HACM114, Deville, HACM360} developed for PDEs, among many others. It has been shown in \cite{HACM259} that analyzing spatial resolution alone is insufficient for solving space-time dependent PDEs while using an alternative to the von Neumann method \cite{LA657} for quantifying numerical stability. Using the global spectral analysis (GSA) developed for space-time dependent problems, the necessary condition for accuracy is that the scheme must preserve the physical dispersion relation obtained from the governing equation and boundary conditions. This is equivalent to stating that for a model 1D convection equation, the numerical group velocity must match with the corresponding physical group velocity for a large range of wavenumbers, as the energy propagates with this parameter \cite{B68, HACM329}. Numerical methods satisfying such criteria are referred to as dispersion relation preserving (DRP) schemes \cite{HACM, HACM312}. Recently, this concept has been used to develop a novel compact scheme in a parallel framework \cite{TOPC} which removes errors due to parallelization using GSA. The added motivation of developing an analysis that did not only comment on the numerical stability of a method \cite{LA657, Charney, Morton} was based on the observation that the solution must propagate at the correct speed with attenuation and dispersion as described by the governing equation. This was found to be all the more important in numerical weather prediction \cite{HACM100}, where the correct propagation of disturbances is vital. The need for developing GSA has also been discussed in \cite{HACM, HACM114,HACM362}.

Originally, compact schemes have been developed for uniform grids. For their application for curved geometries, grid transformation metrics between the physical coordinates and transformed plane coordinates are necessitated \cite{HACM, HACM340,A1}. This mapping involves the use of the chain rule of differentiation to get the terms of grid metrics. As these terms are multiplied with transformed plane uniform-grid derivatives, a major source of aliasing error is introduced \cite{HACM}. For refined computations, using transformation grid metrics slows down the solver significantly. Additionally, this transformation leads to large errors in regions where grid-spacing varies non-smoothly \cite{Gamet}. If one could operate within the physical plane, even while solving for non-uniformly spaced grids, these twin numerical issues of cumbersome metric computations and aliasing errors could be obviated. This is indeed the case for compact schemes developed for non-uniform grids \cite{B96,B86} using the principle of GSA. The scheme developed in \cite{B96}, referred to as the NUC6 scheme, is a sixth-order accurate compact scheme for non-uniform grids. Performance gains were noted for geometries with grids that vary along mutually orthogonal directions using geometric progressions (GP) and tangent hyperbolic stretching functions.

Early efforts to develop compact schemes for non-uniform grids can be attributed to a fourth order compact scheme \cite{Gamet}. Vasilyev \cite{Vasilyev} produced a high order compact scheme on non-uniform grids, from a stencil developed for a uniform grid. Fan \cite{Fan} introduced standard upwind compact schemes which were based completely on upwind-biased stencils for use in non-uniform grids. The authors in \cite{B86} proposed a compact scheme on an arbitrary non-uniform structured grid where the unknown first derivatives are related by a tridiagonal matrix, while the functions involved on the right hand side come from three points. This is referred to as ADB3 scheme, which is formally second order accurate \cite{B86}. Here, we intend to demonstrate the application of compact schemes for spatial discretization on uniform as well as for non-uniform grids with a unique mode of temporal derivative evaluation. This is shown by using the Lax-Wendroff method for obtaining the temporal derivative of model 1D CDE, which offers a novel way of using the governing equation by converting all temporal derivatives into spatial ones. The regions of operation for resolving flow scales in DNS and LES using GSA for the Lax-Wendroff method are quantified.

The Lax-Wendroff method \cite{HACM151} has been applied to a variety of problems due to its ability in preserving the dispersion relation indirectly for standard compact finite differencing techniques \cite{Hirsh2005}. Compact analogs of the Lax-Wendroff method were developed \cite{Hirsh2005}, offering fourth-order accuracy in solving the 1D convection equation. It has also been employed as a correction for wave propagation in attenuating and dispersive media \cite{Blanch}. The method has been used for treating boundaries for compact schemes \cite{Vilar, Tan}. The authors in \cite{Zorio} implement a Lax-Wendroff type procedure to solve hyperbolic equations that allows one to compute time derivatives recursively by using high-order centered differentiation formulae in the Taylor series expansion in time. The same approach has been used in Lax-Wendroff methods for solving hyperbolic PDEs \cite{Winnicki}. We have explored the properties and numerical performance of the Lax-Wendroff method for sixth order compact scheme of Lele \cite{HACM153} for uniform grids and the sixth order NUC6 scheme \cite{B96} for non-uniform grids and quantified the resolution for DNS and LES with prescribed tolerance. A thorough analysis of the resulting numerical schemes is performed using GSA. An estimate of the numerical parameters to be chosen to ensure that spatio-temporal scales are resolved to the accuracy desired in performing DNS and LES is given. This has not been attempted before and will serve as a useful tool for practitioners of HPC. 

The paper is formatted as follows. In Section 2, the Lax-Wendroff method is described from first principle using Taylor series expansion. The extension of Lax-Wendroff method for solving the 1D CDE is shown in Section 3. The analysis of the 1D CDE using GSA is described in Section 4. In Section 5, we describe a novel implementation of the Lax-Wendroff method for non-uniform grids using NUC6 scheme. The numerical performance of the ensuing numerical method is characterized using the model 1D CDE. In Section 6, the Lax-Wendroff method in the previous section is applied for a uniform grid and analyzed using GSA. The choice of numerical parameters to ensure the DRP property of the newly developed methods is quantified for resolving scales in DNS and LES in Section 7. The paper closes with a summary and concluding remarks.

\section{Lax-Wendroff Method}
\label{LWS}
In the Lax-Wendroff method, second order accuracy for time integration is obtained using the Taylor series expansion, 

\begin{equation}
u(t+\Delta t)= u(t) + \Delta t \frac{\partial u}{\partial t} + \frac{(\Delta t)^2}{2} \frac{\partial^2 u}{\partial t^2}
\label{LWS1}
\end{equation}

For one-dimensional (1D) convection equation, the first time derivative is obtained from the governing equation,
$$\frac{\partial u}{\partial t} = -c \frac{\partial u}{\partial x}$$
\noindent and the second order derivative is obtained by differentiating the above equation with respect to time to obtain,
$$\frac{\partial^2 u}{\partial t^2} = c^2 \frac{\partial^2 u}{\partial x^2}$$
Substituting the first and second time derivatives in Eq. \eqref{LWS1}, one obtains the modified differential equation (MDE) as,

\begin{equation}
u(t + \Delta t)= u(t) -c \Delta t \frac{\partial u}{\partial x} + \frac{c^2 (\Delta t)^2}{2} \frac{\partial^2 u}{\partial x^2}
\label{LWS2}
\end{equation}

The Lax-Wendroff method provides second order time-accuracy by a two-time level method without spurious modes, while satisfying the governing equation. One notes the presence of an additional diffusive term in MDE which is not present in the original governing equation. This is an example of a hyperbolic partial differential equation being transformed into a convection-diffusion equation (CDE). This added diffusive term is stabilizing, but converting the hyperbolic differential equation to a parabolic differential equation shows the lack of consistency. The converse is noted for the Dufort-Frankel method \cite{HACM,HACM65}, where the heat equation is solved as a wave propagation problem for specific numerical parameters. 

\section{Lax-Wendroff Method for 1D Convection-Diffusion Equation}

The CDE is a canonical linearized model of the Navier-Stokes equation, as has been established in \cite{B118} for compact schemes. Unlike the pure convection equation, the Lax-Wendroff method for CDE maintains the diffusive nature and the problem of inconsistency is absent even though a third derivative dispersive term makes its appearance along with the higher order diffusion term. The effects of additional terms on the accuracy of space-time discretization are studied. The governing CDE is written for easy reference as,

\begin{equation}
\frac{\partial u}{\partial t} + c\frac{\partial u}{\partial x} = \alpha \frac{\partial^2u}{\partial x^2}
\label{eqn22}
\end{equation}

For the Lax-Wendroff method, substitution in Eq. \eqref{LWS1} yields the following,

\begin{equation}
u(t + \Delta t)= u(t) -c \Delta t \frac{\partial u}{\partial x} + \alpha \Delta t \frac{\partial^2 u}{\partial x^2} + \frac{(\Delta t)^2}{2} \frac{\partial^2 u}{\partial t^2}
\label{eqn23}
\end{equation}

From Eq. \eqref{eqn22} one obtains

\begin{equation}
 \frac{\partial^2 u}{\partial t^2} = c^2 \frac{\partial^2 u}{\partial x^2} - 2 \alpha c \frac{\partial^3 u}{\partial x^3} + \alpha^2 \frac{\partial^4 u}{\partial x^4}
\label{eqn24}
\end{equation}

Substitution of Eq. \eqref{eqn24} in Eq. \eqref{eqn23} results in added diffusive terms from the convection and diffusion terms. The dispersive term is due to the interaction between convection and diffusion terms. A fourth order diffusion term appears from the physical diffusion term upon the application of the Lax-Wendroff method. From Eq. \eqref{eqn24}, it is clear that a numerical method when used to solve the 1D CDE, invokes the diffusion differently depending on the length scale only. With the Lax-Wendroff method, this amounts to analyzing the spatial discretization only, which is a unique feature of this method. The retention of the higher order terms for time integration in the Taylor series will involve third and fourth order terms in spatial discretization, which need to be addressed separately. We will describe this in Sec. 5, in context of compact scheme for non-uniformly spaced grids. 

\section{Global Spectral Analysis: 1D Convection-Diffusion Equation with Lax Wendroff Method}

In this section, we describe in brief the GSA for 1D CDE which is then used to analyze the Lax-Wendroff method. These are described in \cite{B94}, and recounted here for the ease of understanding. For the 1D CDE, with $c$ and $\alpha$ as real constants, one substitutes the hybrid spectral representation, $u(x,t) = \int \hat{U}(k,t)e^{ikx}dk$ to obtain,

\begin{equation}
\frac{d\hat{U}}{dt} + ick\hat{U} = -\alpha k^{2}\hat{U}.
\label{eqn26}
\end{equation}

\noindent which for a general initial condition: $u(x_j,t = 0) = u_j^0 = \int U_0 (k)\; e^{ikx_j} dk$ can be solved analytically to obtain,
\begin{equation}
 \hat{U}(k,t) = U_0(k) \: e^{-\alpha k^{2}t} e^{-ikct}.
 \label{eqn27}
\end{equation}

Employing the bilateral transform, $u(x,t) = \int \int U(k, \omega_0) e^{i(kx - \omega_0 t)} dk\; d\omega_0$, the physical dispersion relation for the CDE is obtained as,

\begin{equation}
 \omega_0 = c \: k - i \:\alpha \: k^{2}.
 \label{eqn28}
\end{equation}

Equation \eqref{eqn28} yields the physical phase speeds as,
\begin{equation}
  c_{phys} = \frac{\omega_0}{k} =  c - i \: \alpha k,
  \label{eqn29}
\end{equation}

The real part of the phase speed is taken as it is the common practice in the literature. From the dispersion relation, the physical group velocity (energy propagation velocity of a wave-packet), is obtained following the definition given in \cite{HACM, HACM329},

\begin{equation}
  V_{g,phys} = \frac{d\omega_0}{dk} = c - 2 \: i \: \alpha k.
  \label{eqn30}
\end{equation}

From the above relation, the group velocity is noted as a complex quantity. Its physical implication is given in \cite{B94} by noting that the coefficient of diffusion is real and that leads to the real part of the group velocity as $c$.

The analytical solution of CDE in Eq. \eqref{eqn27} helps identify the physical amplification factor as the ratio of the solution amplitude at two distinct time instants separated by $\Delta t$, given by 

\begin{equation}
  G_{phys} = \frac{\hat{U}(k,t+\Delta t)}{\hat{U}(k,t)} =
  e^{-\alpha \: k^{2}\Delta t} e^{-i \: k \: c \: \Delta t} = e^{-i \: \omega_0\: \Delta t }
  \label{eqn31}
\end{equation}

This is expressed in terms of the numerical parameters, CFL ($N_c$) and Peclet ($Pe$) numbers as,

\begin{equation}
  G_{phys} = e^{-Pe \: (kh)^{2}} \: e^{-i \: N_c \: (kh)}.
  \label{eqn32}
\end{equation}

The primary distinction between GSA and the popular von-Neumann analysis is noted at this stage. Contrary to the popular assumption following von-Neumann analysis, it is noted that the numerical phase speed ($c_{num}$), and the numerical diffusion ($\alpha_{num}$), are dependent on $k$, $h$, $N_c$ and $Pe$ \cite{B94, B118, B119}. Similar to the physical solution, the numerical method has a numerical dispersion relation and an associated numerical amplification factor, $G_{num}$, which dictates the evolution of the solution. Thus, it is necessary that for the numerical method, $G_{num}$ must be as close to $G_{phys}$ as possible, for an accurate solution.

The numerical dispersion relation is analogous to Eq.~(\ref{eqn28}), which is expressed as \cite{B94},

\begin{equation}
  \omega_{num} = c_{num} \: k - i \: \alpha_{num} \: k^{2}.
  \label{eqn33}
\end{equation}

Here, $\omega_{num}$ is complex and differs from the physical dispersion relation expression, as $c_{num}$ and $\alpha_{num}$ vary with $kh$, $N_c$ and $Pe$. Noting the numerical dispersion relation, the numerical amplification factor ($G_{num}$) can be obtained as a function of $\alpha_{num}$ and $c_{num}$ as,

\begin{equation}
  G_{num} = e^{-\alpha_{num} \: k^{2}\Delta t} \: e^{-i \: k \:c_{num} \Delta t} = e^{-i \: \omega_{num} \Delta t}.
  \label{eqn34}
\end{equation}

From the complex $G_{num}$, the phase shift per time step is evaluated to obtain $c_{num}$. Following similar previous relations it is noted,

\begin{equation}
  c_{num} = \Re\left(\frac{\omega_{num}}{k}\right),   \label{eqn35}
\end{equation}
and,
\begin{equation}
   V_{g,num} = \Re\left(\frac{d\omega_{num}}{dk}\right).
   \label{eqn36}
\end{equation}

The numerical phase shift per time step ($\beta$) is evaluated as the ratio between the imaginary and real part of $G_{num}$ given by,

\begin{equation}
  tan(\beta) = -\frac{\Im(G_{num})} {\Re(G_{num})}
  =  tan(c_{num} k \Delta t).
\label{eqn37}
\end{equation}
\noindent where the $\Re$ and $\Im$ denote real and imaginary parts, respectively. This is also the phase shift per time step used before as $\phi_j$ for the $j^{th}$
node. From this, the non-dimensional numerical phase speed is obtained as,

\begin{equation}
  \frac{c_{num}}{c_{phys}} = \frac{\beta}{kc\Delta t} =
-\frac{1}{(kh)N_c} \: tan^{-1}\left[\frac{\Im(G_{num})} {\Re(G_{num})}\right].
 \label{eqn38}
\end{equation}

From $c_{num}$, the numerical group velocity is obtained as,

\begin{equation}
V_{g,num} = \Re\left( \frac{d \omega_{num}}{dk}\right) = \frac{1}{\Delta t}{d \beta \over dk},
\label{eqn39}
\end{equation}

\noindent which when expressed in non-dimensional form one gets,

\begin{equation}
 \frac {V_{g,num}} {V_{g,phys}} = \frac{1}{N_c}\frac{d\beta}{d(kh)}.
 \label{eqn40}
\end{equation}

From Eq.~\eqref{eqn32}, it is noted that the numerical amplification factor modulus is only dependent on $\alpha_{num}$ as,

\begin{equation}
   |G_{num}| = e^{-\alpha_{num} k^{2}\Delta t},
   \label{eqn41}
\end{equation}

This can be expressed in terms of $Pe$ as,

\begin{equation}
    ln|G_{num}| = -\frac{\alpha_{num}} {\alpha} (kh)^{2} \: Pe.
    \label{eqn42}
\end{equation}

Equation~\eqref{eqn42} is used to evaluate the numerical diffusion coefficient in non-dimensional form as,

\begin{equation}
    \frac {\alpha_{num}}{\alpha} = -\frac {ln|G_{num}|}{(kh)^{2} Pe}.
    \label{eqn43}
\end{equation}

\section{Application of Lax-Wendroff Method on Non-Uniform Grids}
Here, we develop a new Lax-Wendroff method that employs non-uniform structured grids for a compact scheme, with the purpose of computing in the physical plane \cite{B96}. Additionally, we require two sixth-order non-uniform compact schemes for the second and fourth order spatial derivatives appearing in Eq. \ref{eqn24}, for the Lax-Wendroff method. The third derivative is also needed, which will follow directly from the other derivatives. The NUC6 scheme developed in \cite{B96} for the first derivative using a non-uniform grid in the physical plane provides the basis for these new symmetric schemes for higher even derivatives. The third derivative required by the Lax-Wendroff method in Eq. \eqref{eqn24} is obtained by a tensor product involving first and second derivatives, as described later in this section. The rationale for adopting a tensor product (instead of a separate compact scheme directly) for the third derivative, is due to the highly dispersive nature of the symmetric compact scheme developed based on the NUC6 scheme \cite{B96}. Such a compact scheme for the third derivative causes Gibbs oscillations in both upstream and downstream directions. In contrast, the proposed tensor product based scheme for the third derivative does not exhibit spurious dispersion, as the first derivative and non-dispersive second derivative discretizations are not a source of high dispersion.

As noted before, two sixth order compact stencils have to be developed for $2^{nd}$ and $4^{th}$ order derivative terms to solve the 1D CDE using the Lax-Wendroff method in Eq. \eqref{eqn24}. The rationale for such a choice is due to the proven accuracy of NUC6 scheme \cite{B96}, and also because the stencils for $1^{st}$, $2^{nd}$ and $4^{th}$ order derivatives on uniform grids reduce to the well known sixth order interior stencils of schemes proposed by Lele \cite{HACM153}.
For a general non-uniform structured grid in 1D with $M$ points, the stencils for computing the first, second and fourth derivatives are given as:

\begin{equation}
\begin{split}
&\text{First Derivative:}\\
&u^{\prime}_1=\frac{1}{h_{r2}}\left[\left(-(\beta_2+1)+\frac{1}{\beta_2+1} \right)u_1 + (\beta_2+1)u_2 -\frac{u_3}{\beta_2+1}\right]\\[1.5ex]
&\frac{u^{\prime}_1}{2} + u^{\prime}_2 + \frac{u^{\prime}_3}{2} = \frac{u_2-u_1}{h_{l2}} + \frac{u_3-u_2}{h_{r2}}\\[1.5ex]
&p^1_{j-1}u^{\prime}_{j-1} + p^1_{j}u^{\prime}_{j} + p^1_{j+1}u^{\prime}_{j+1}=s^1_{-2}(u_j-u_{j-2}) + s^1_{-1}(u_j-u_{j-1})+ s^1_{1}(u_{j+1}-u_j)\\&\hspace{0mm}+ s^1_{2}(u_{j+2}-u_j)\\[1.5ex]
&\frac{u^{\prime}_{M-2}}{2} + u^{\prime}_{M-1} + \frac{u^{\prime}_M}{2} = \frac{u_{M-1}-u_{M-2}}{h_{lM-1}} + \frac{u_M-u_{M-1}}{h_{rM-1}}\\[1.5ex]
&u^{\prime}_M=\frac{1}{h_{rM-1}}\left[\left((\beta_{M-1}+1)-\frac{\beta^2_{M-1}}{\beta_{M-1}+1} \right)u_M - (\beta_{M-1}+1)u_{M-1} +\frac{\beta^2_{M-1} u_{M-2}}{\beta_{M-1}+1}\right]\\
\end{split}
\label{der1}
\end{equation}

\begin{equation}
\begin{split}
&\text{Second Derivative:}\\
\hspace{1mm}&u^{\prime \prime}_1=\frac{2(h_{r1}+h_{rr1}+h_{rrr1})}{h_{r1}h_{rr1}h_{rrr1}}u_1 - \frac{2(h_{rr1}+h_{rrr1})}{h_{r1}(h_{r1}-h_{rrr1})(h_{r1}-h_{rr1})}u_2\\
\hspace{1mm}&\hspace{10mm} - \frac{2(h_{r1}+h_{rrr1})}{h_{rr1}(h_{r1}-h_{rr1})(h_{rrr1}-h_{rr1})}u_3 + \frac{2(h_{r1}+h_{rr1})}{h_{rrr1}(h_{r1}-h_{rrr1})(h_{rrr1}-h_{rr1})}u_4\\[1.5ex]
\hspace{1mm}&u^{\prime\prime}_2 = \frac{2}{h_{l2}(h_{l2}+h_{r2})}u_1 - \frac{2}{h_{l2}h_{r2}}u_2 + \frac{2}{h_{r2}(h_{l2}+h_{r2})}u_3\\[1.5ex]
\hspace{1mm}&p^2_{j-1}u^{\prime \prime}_{j-1} + p^2_{j}u^{\prime \prime}_{j} + p^2_{j+1}u^{\prime \prime}_{j+1}=s^2_{-2}u_{j-2} + s^2_{-1}u_{j-1}+ s^2_0 u_j + s^2_{1}u_{j+1}+ s^2_{2}u_{j+2}\\[1.5ex]
\hspace{1mm}&u^{\prime \prime}_{M-1} = \frac{2}{h_{lM-1}(h_{lM-1}+h_{rM-1})}u_{M-2} - \frac{2}{h_{lM-1}h_{rM-1}}u_{M-1} + \frac{2}{h_{rM-1}(h_{lM-1}+h_{rM-1})}u_M\\[1.5ex]
\hspace{1mm}&u^{\prime \prime}_M=\frac{2(h_{lM}+h_{llM})}{h_{lllM}(h_{lM}-h_{lllM})(h_{lllM}-h_{llM})}u_{M-3} - \frac{2(h_{lM}+h_{lllM})}{h_{llM}(h_{lM}-h_{llM})(h_{lllM}-h_{llM})}u_{M-2} \\
\hspace{1mm}&\hspace{10mm} - \frac{2(h_{lllM}+h_{llM})}{h_{lM}(h_{lM}-h_{lllM})(h_{lM}-h_{llM})}u_{M-1}+ \frac{2(h_{lM}+h_{llM}+h_{lllM})}{h_{lM}h_{llM}h_{lllM}}u_M\\
\end{split}
\label{der2}
\end{equation}

\begin{equation}
\begin{split}
&\text{Fourth Derivative:}\\
&u^{\prime \prime \prime \prime}_1 = \frac{24}{h_{r1}h_{rr1}h_{rrr1}h_{rrrr1}}u_1 + \frac{24}{h_{r1}(h_{r1}-h_{rrr1})(h_{r1}-h_{rrrr1})(h_{r1}-h_{rr1})}u_2 \\
&\hspace{10mm}-\frac{24}{h_{rr1}(h_{r1}-h_{rr1})(h_{rrr1}-h_{rr1})(h_{rrrr1}-h_{rr1})}u_3\\
&\hspace{10mm}-\frac{24}{h_{rrr1}(h_{r1}-h_{rrr1})(h_{rrr1}-h_{rrrr1})(h_{rrr1}-h_{rr1})}u_4 \\
&\hspace{10mm}+\frac{24}{h_{rrrr1}(h_{r1}-h_{rrrr1})(h_{rrr1}-h_{rrrr1})(h_{rrrr1}-h_{rr1})}u_5\\[1.5ex]
&u^{\prime \prime \prime \prime}_2 = \frac{24}{h_{l2}(h_{l2} + h_{r2})(h_{l2}+h_{rrr2})(h_{l2}+h_{rr2})}u_1 - \frac{24}{h_{l2}h_{r2}h_{rr2}h_{rrr2}}u_2 \\
&\hspace{10mm}+\frac{24}{h_{r2}(h_{l2}+h_{r2})(h_{r2}-h_{rrr2})(h_{r2}-h_{rr2})}u_3 + \frac{24}{h_{rr2}(h_{l2}+h_{rr2})(h_{r2}-h_{rr2})(h_{rrr2}-h_{rr2})}u_4 \\
&\hspace{10mm}- \frac{24}{h_{rrr}(h_{l}+h_{rrr})(h_{r}-h_{rrr})(h_{rrr}-h_{rr})}u_5\\[1.5ex]
&p^4_{j-1}u^{\prime \prime \prime \prime}_{j-1} + p^4_{j}u^{\prime \prime \prime \prime}_{j} + p^4_{j+1}u^{\prime \prime \prime \prime}_{j+1}=s^4_{-2}u_{j-2} + s^4_{-1}u_{j-1}+ s^4_0 u_j + s^4_{1}u_{j+1}+ s^4_{2}u_{j+2}\\[1.5ex]
&u^{\prime \prime \prime \prime}_{M-1} = -\frac{24}{h_{lllM-1}(h_{lllM-1}+h_{rM-1})(h_{lM-1}-h_{lllM-1})(h_{lllM-1}-h_{llM-1})}u_{M-4} \\
&\hspace{10mm}+ \frac{24}{h_{llM-1}(h_{llM-1}+h_{rM-1})(h_{lM-1}-h_{llM-1})(h_{lllM-1}-h_{llM-1})}u_{M-3}  \\
&\hspace{10mm}+\frac{24}{h_{lM-1}(h_{lM-1}+h_{rM-1})(h_{lM-1}-h_{lllM-1})(h_{lM-1}-h_{llM-1})}u_{M-2}\\
&\hspace{10mm}-\frac{24}{h_{lM-1}h_{llM-1}h_{lllM-1}h_{rM-1}}u_{M-1}\\
&\hspace{10mm}+ \frac{24}{h_{r}(h_l + h_r)(h_{lll}+h_r)(h_{ll}+h_r)}u_M\\[1.5ex]
&u^{\prime \prime \prime \prime}_M = \frac{24}{h_{llllM}(h_{lM}-h_{llllM})(h_{lllM}-h_{llllM})(h_{llllM}-h_{llM})}u_{M-4}\\
&\hspace{10mm}-\frac{24}{h_{lllM}(h_{lM}-h_{lllM})(h_{lllM}-h_{llllM})(h_{lllM}-h_{llM})}u_{M-3}\\
&\hspace{10mm}-\frac{24}{h_{llM}(h_{lM}-h_{llM})(h_{lllM}-h_{llM})(h_{llllM}-h_{llM})}u_{M-2}\\
&\hspace{10mm}+\frac{24}{h_{lM}(h_{lM}-h_{lllM})(h_{lM}-h_{llllM})(h_{lM}-h_{llM})}u_{M-1}  \\
&\hspace{10mm}+\frac{24}{h_{lM}h_{llM}h_{lllM}h_{llllM}}u_M \\
\end{split}
\label{der4}
\end{equation}

\noindent where the variables used are given by, $h_{li}=x_i-x_{i-1}$, $h_{lli}=x_i-x_{i-2}$, $h_{llli}=x_i-x_{i-3}$, $h_{lllli}=x_i-x_{i-4}$, $h_{ri}=x_{i+1}-x_{i}$, $h_{rri}=x_{i+2}-x_{i}$, $h_{rrri}=x_{i+3}-x_{i}$, $h_{rrrri}=x_{i+4}-x_{i}$, respectively. Variables $\beta_2$ and $\beta_{M-1}$ denote the stretching ratios and are defined as $\beta_2=h_{r2}/h_{r1}$, $\beta_{M-1}=h_{rM-1}/h_{lM-1}$. The Taylor series coefficients of various orders are matched to obtain a linear relation between $p^1_{j-1}$, $p^1_{j+1}$, $p^2_{j-1}$, $p^2_{j+1}$, $p^4_{j-1}$, $p^4_{j+1}$, $s^1_{-2}$, $s^1_{-1}$, $s^1_{0}$, $s^1_{1}$, $s^1_{2}$, $s^2_{-2}$, $s^2_{-1}$, $s^2_{0}$, $s^2_{1}$, $s^2_{2}$, $s^4_{-2}$, $s^4_{-1}$, $s^4_{0}$, $s^4_{1}$ and $s^4_{2}$. The linear system of equations is solved using MATLAB\textsuperscript{\textregistered} Symbolic Math Toolbox\textsuperscript{\texttrademark} and the expression for these coefficients are provided in Appendix.

The compact schemes for the full domain can be represented in the general matrix-vector form in the physical plane as,

\begin{equation}
\begin{split}
&[{\bf A}_1]\{{\bf u}^{\prime}\} = [{\bf B}_1]\{{\bf u}\}\\
&[{\bf A}_2]\{{\bf u}^{\prime\prime}\} = [{\bf B}_2]\{{\bf u}\}\\
&[{\bf A}_4]\{{\bf u}^{\prime\prime\prime\prime}\} = [{\bf B}_4]\{{\bf u}\}
\end{split}
\label{mateq}
\end{equation}

This also explains why the non-uniform compact scheme is preferred, as the equivalent form of the CDE given by Eq. \eqref{eqn24} in the transformed plane will have many additional contributions arising from grid metric transformation terms \cite{HACM, A1}. In the present formalism, one only needs to cater to the terms given in Eq. \eqref{eqn24}.  

One can also represent the derivatives given above in linear algebraic form by their equivalent explicit expressions as,

\begin{equation}
\begin{split}
&\{{\bf u}^{\prime}\} = [{\bf C}_1]\{{\bf u}\}\\
&\{{\bf u}^{\prime\prime}\} = [{\bf C}_2]\{{\bf u}\}\\
&\{{\bf u}^{\prime\prime\prime\prime}\} = [{\bf C}_4]\{{\bf u}\}
\end{split}
\label{inv_mateq}
\end{equation}

\noindent where the matrices on the right hand sides of the above equations are given as, $[{\bf C}_1]=[{\bf A}_1]^{-1}[{\bf B}_1]$, $[{\bf C}_2]=[{\bf A}_2]^{-1}[{\bf B}_2]$ and $[{\bf C}_4]=[{\bf A}_4]^{-1}[{\bf B}_4]$.

The $3^{rd}$ derivative required by the Lax-Wendroff method is computed specifically by a tensor operation given by

\begin{equation}
\{{\bf u}^{\prime\prime\prime}\} = [{\bf C}_1][{\bf C}_2]\{{\bf u}\}
\label{der3}
\end{equation}

We note that in actual application, the matrices $[{\bf C}_1]$ to $[{\bf C}_4]$ are not computed. Instead, the well known, fast and accurate Thomas algorithm is used for computing the derivatives due to tridiagonal structure of the $[{\bf A}_1]$ to $[{\bf A}_4]$ matrices.

Using the representation for the derivatives given in Eqs. \eqref{inv_mateq} and \eqref{der3}, the numerical amplification factor for the
scheme can be determined as,

\begin{equation}
 G_j = 1 - \frac{N_c}{2} \sum_{l=1}^M C_{1,jl}P_{lj} +(Pe +\frac{N_c^2}{2}) \sum_{l=1}^M C_{2,jl}P_{lj} - (Pe N_c) \sum_{l=1}^M C_{3,jl}P_{lj} + \frac{Pe^2}{2} \sum_{l=1}^M C_{4,jl}P_{lj}
\label{gn_nuc6}
\end{equation}

\noindent where, $j$ denotes the node for which $G$ is evaluated, $C_{(),jl}$ refers to the $(j,l)^{th}$ entry of the corresponding matrix and
$l$ is the nodal index which is summed from $1$ to $M$. The quantity $P_{lj}$ is the projection matrix and is defined as $P_{lj}=e^{ik(x_l-x_j)}$ \cite{HACM}.

For the non-uniform grid, a suitable length scale must be chosen to define the nondimensional parameters $N_c$ and $Pe$ in Eq. \eqref{gn_nuc6}, as well as non-dimensionalize the wavenumber ($k$), so that these numerical properties can be compared correctly across different nodes. In this regard, we employ the choice proposed in \cite{B86}, where the authors proposed the non-uniform compact scheme. Hence in the following analysis, the length scale is chosen as $h_{rj}$, an appropriate choice that incorporates the effect of variation in spacing across the nodes.

In the present study based on non-uniform grid compact schemes, the analysis for the Lax-Wendroff method is performed for a grid generated by the tangent-hyperbolic function. The tangent hyperbolic distribution is a popular choice for generating non-uniform structured grids, and is well known for its anti-aliasing property \cite{HACM, A1}. In this approach, one generates a non-uniform grid given by,

\begin{equation}
x_j=L\left[1-{\tanh\gamma(1-\eta_j)\over\tanh\gamma}\right];  \quad \quad \text{with } \eta_j={j-1\over M-1} \,\; {\rm and} \;\, j=1,2,\cdots,M-1,M
\label{tanhgrid}
\end{equation}

\noindent where $\gamma$ is the parameter controlling the spacing, $M$ is the total number of points of the grid and $L$ is the size of the domain. Increasing $\gamma$ value results in more stretching of the grid in the left direction. For evaluating the property charts of the numerical scheme, we choose a grid such that the domain length is $L=1$ and the total number of points is $M =201$.

    \begin{figure}[H]
        \centering
        \includegraphics[scale=0.55]{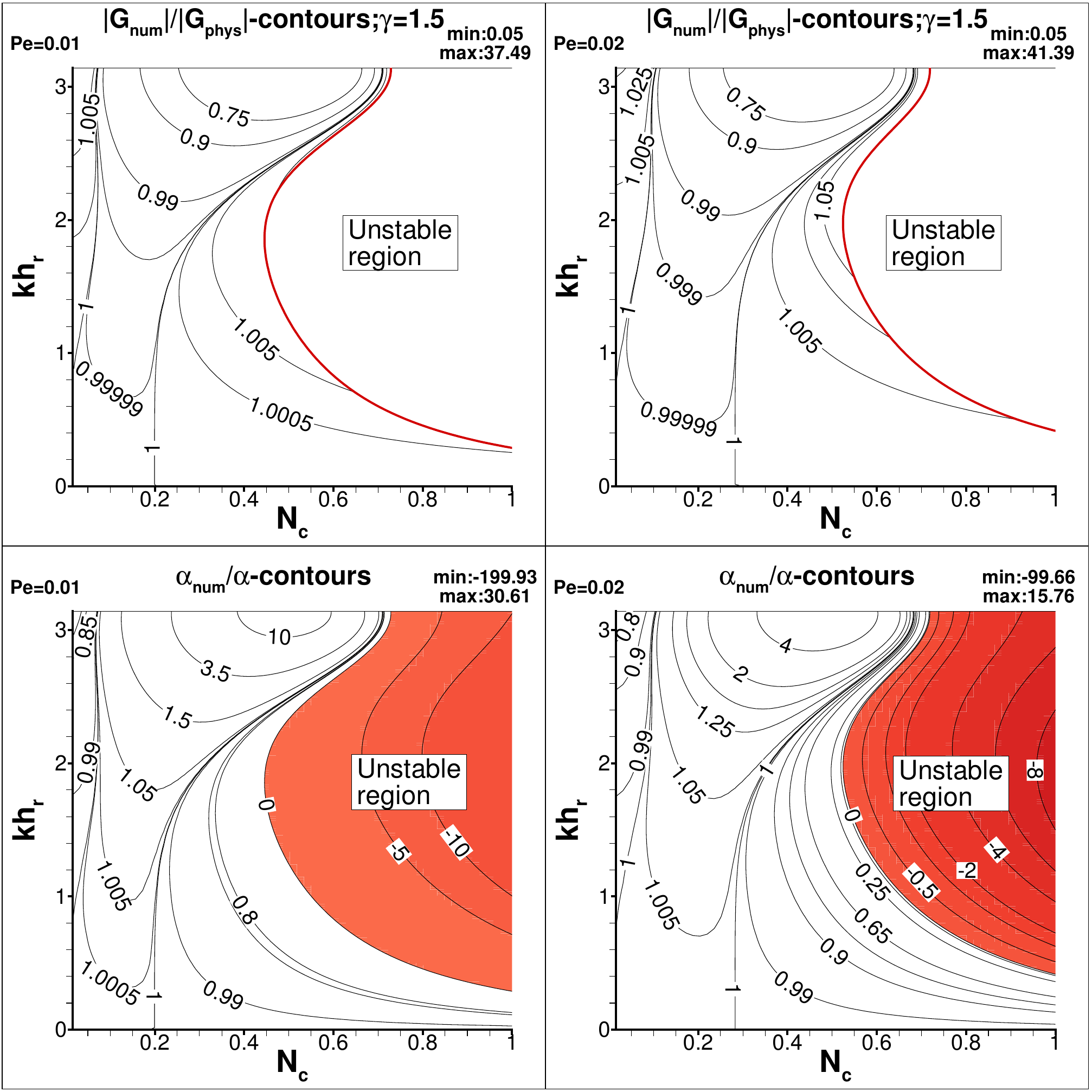}
        \caption{Property charts of the full Lax-Wendroff method using a non-uniform compact scheme for the CDE plotted in the $(N_c,kh_r)$-plane for the Peclet numbers $Pe=0.01$ and $0.02$ for a tangent hyperbolic grid with $\gamma=1.5$. The ratio of numerical and physical amplification factors $\left(\frac{|G_{num}|}{|G_{phys}|}\right)$ and the ratio of numerical to the physical diffusion coefficients $\left(\frac{\alpha_{num}}{\alpha}\right)$ are shown. Regions of numerical instability are marked in the panels.}
        \label{fig9}
    \end{figure}

    \begin{figure}[H]
        \centering
         \includegraphics[scale=0.55]{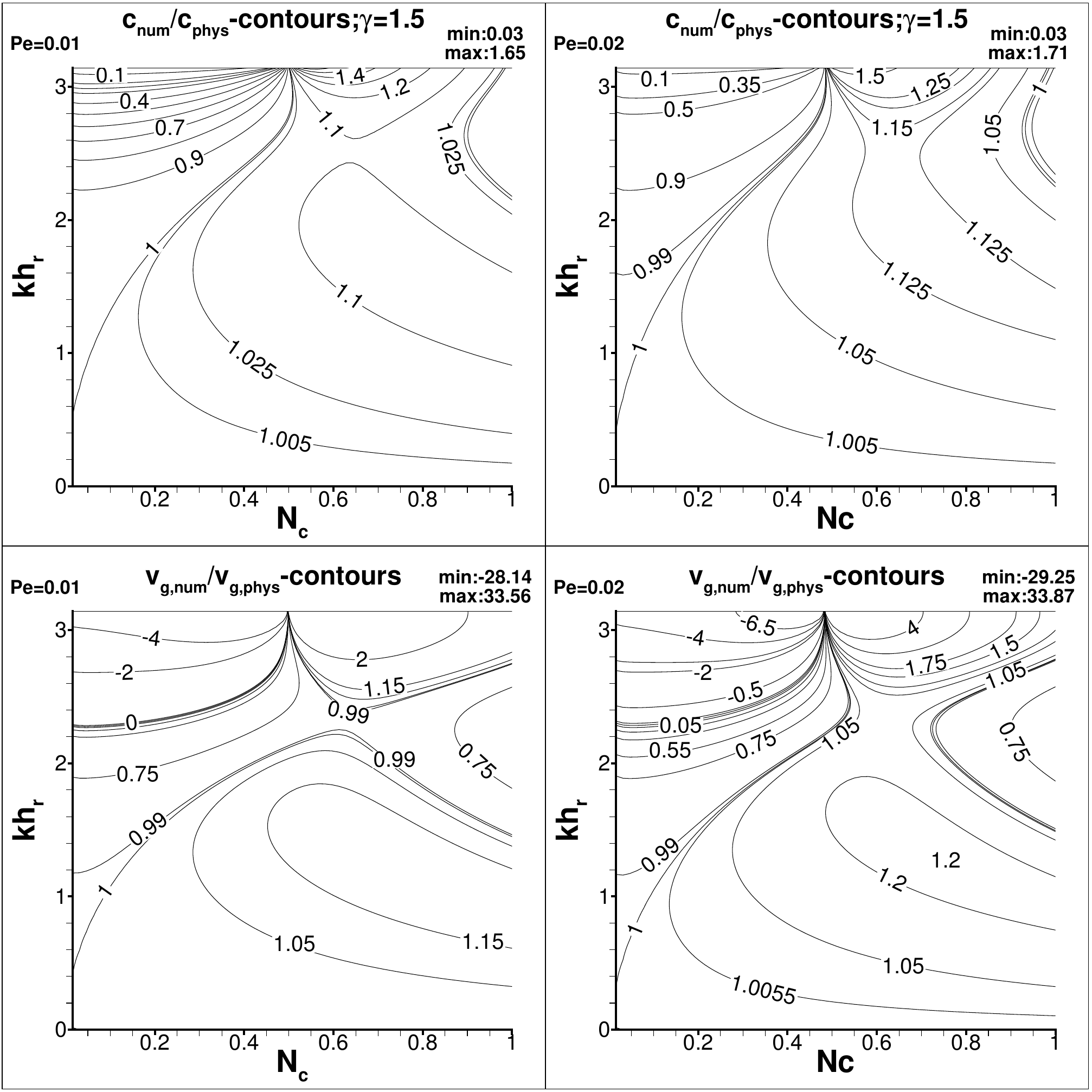}
        \caption{Property charts of the full Lax-Wendroff method using a non-uniform compact scheme for the CDE plotted in the $(N_c,kh_r)$-plane for the Peclet numbers $Pe=0.01$ and $0.02$ for a tangent hyperbolic grid with $\gamma=1.5$. The ratio of numerical and physical physical phase speeds $\frac{c_{num}}{c_{phys}}$ and the ratio of numerical to the physical group velocities $\frac{V_{g,num}}{V_{g,phys}}$ are plotted.}
        \label{fig10}
    \end{figure}

In Figs. \ref{fig9} and \ref{fig10}, property charts are shown in $(N_c,kh_r)$-plane for $Pe=0.01$ and $0.02$, for the Lax-Wendroff method based on non-uniform compact schemes given by Eqns. \eqref{der1}, \eqref{der2}, \eqref{der4} and \eqref{der3} for the CDE. The properties are shown for the stretching parameter $\gamma=1.5$, for the central node $j=101$, of the grid. Plotted quantities show that the numerical scheme has good numerical properties for $\alpha_{num}$, $c_{num}$ and $V_{g, num}$, comparable to the corresponding physical values over a reasonable range of wavenumbers $kh_r$ for $N_c \le 1.0$ indicating good numerical accuracy. One notes that at lower $N_c$, $\alpha_{num}$ becomes progressively lower than 1, whereas $\alpha_{num}$ increases above 1, at all wavenumbers for a range of $N_c$ values, thereby indicating lower and higher than exact physical diffusion, respectively. It is also noted that increasing $Pe$ value results in extension of the region of accuracy along with improvement of accuracy at higher wavenumbers. There are two features which are noteworthy: The $|G_{num}|/|G_{phys}| = 1$ contour intersects the $N_c$-axis at only one point, on the lower range of the $N_c$-axis. Secondly, this particular contour value, for which the numerical amplification factor is coincident with the physical amplification factor, emerges from the $N_c$ axis almost vertically as a straight line. As a consequence, one can draw boxes ABCD and EFGH in which the DRP property of the numerical method is ensured and will be used later for compact schemes (in Figs. \ref{fig15} and \ref{fig16}), which also gives us an estimate of the maximum wavenumber that can be resolved, $(kh)_{max}$. This in turn provides us with an estimate of how many points are required to resolve a wave specified for a required precision, which is referred to as points per wave (PPW). In this way, the property chart can be interpreted for the 1D CDE using non-uniform compact schemes. We note that the admissible value of $N_c$ for the compact schemes will be significantly lower, compared to the Lax-Wendroff method based on the central difference schemes. One also notes that the boxes ABCD and EFGH do not identify $N_c$ ranges which are unstable regions with respect to anti-diffusion (i.e. $\alpha_num/\alpha < 0$). This is one of the strongest points in favour of using compact schemes. The superior dispersion properties of compact schemes are also noted in Fig. \ref{fig10}, with negligible phase and dispersion errors in the identified region, based on lower tolerance values for $|G_{num}|/|G_{phys}|-1$. 

    \begin{figure}[H]
        \centering
         \includegraphics[scale=0.55]{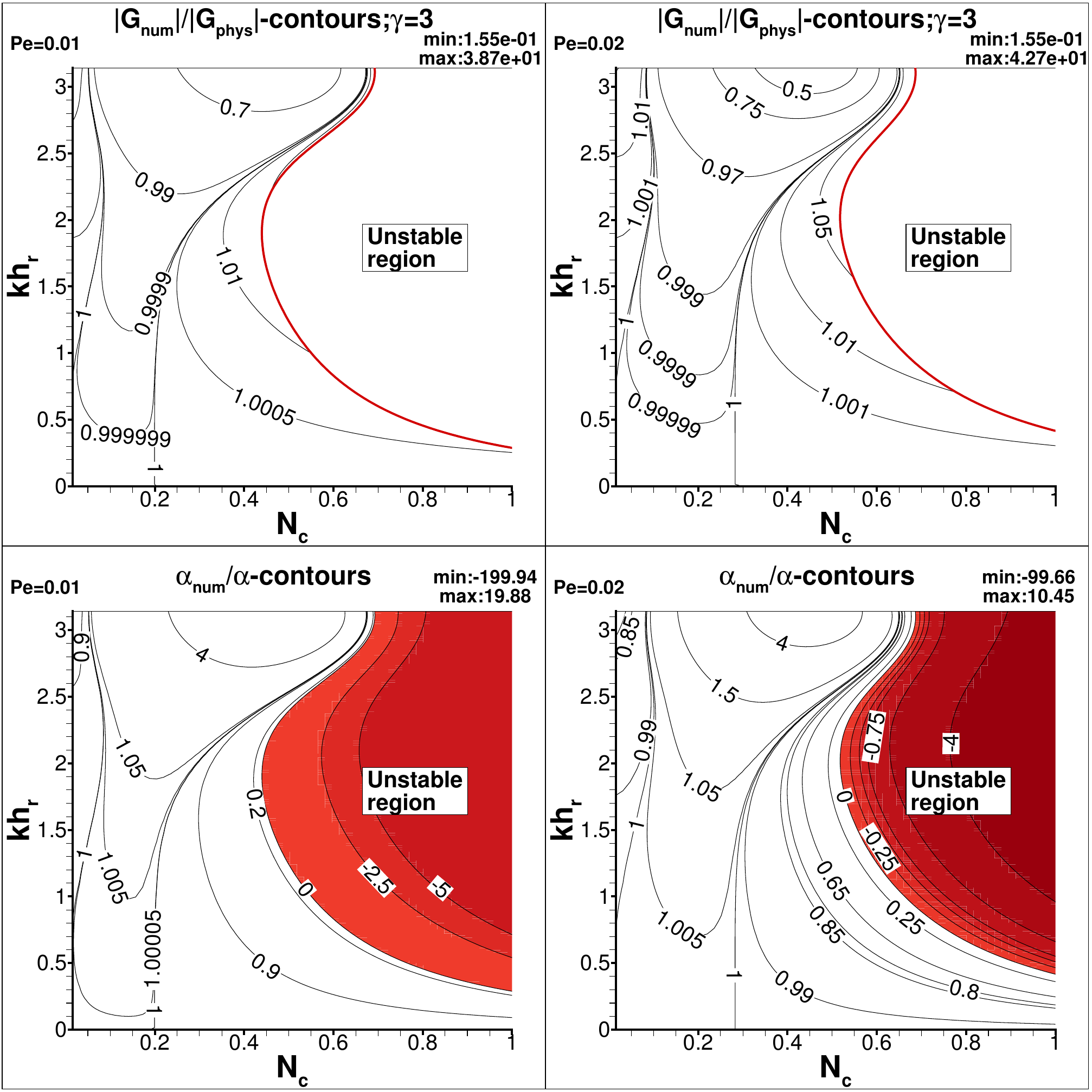}
        \caption{Property charts of the full Lax-Wendroff method using a non-uniform compact scheme for the CDE plotted in the $(N_c,kh_r)$-plane for the Peclet numbers $Pe=0.01$ and $0.02$ for a tangent hyperbolic grid with $\gamma=3$. Plotted are the ratio of numerical to the corresponding physical amplification factors 
$\left(\left|\frac{G_{num}}{G_{phys}}\right|\right)$ and the ratio of numerical to the physical diffusion coefficients $\left(\frac{\alpha_{num}}{\alpha}\right)$. Regions of numerical instability are as marked in the panels.}
        \label{fig11}
    \end{figure}

    \begin{figure}[H]
        \centering
         \includegraphics[scale=0.55]{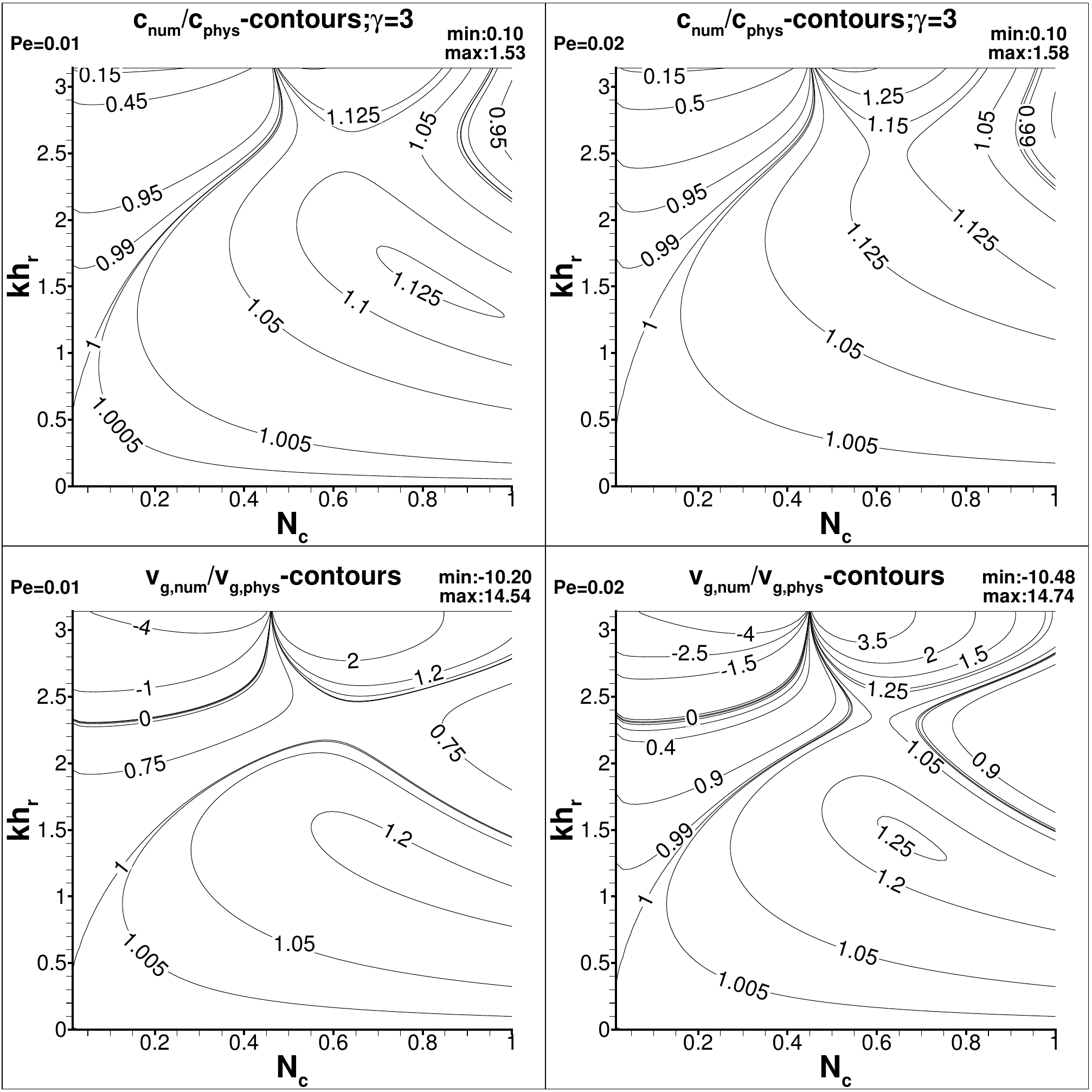}
        \caption{Property charts of the full Lax-Wendroff method using a non-uniform compact scheme for the CDE plotted in the $(N_c,kh_r)$-plane for the Peclet numbers $Pe=0.01$ and $0.02$ for a tangent hyperbolic grid with $\gamma=3$. Plotted are the ratio of numerical to the corresponding physical phase speeds $\frac{c_{num}}{c_{phys}}$ and the ratio of numerical to the physical group velocity $\frac{V_{g,num}}{V_{g,phys}}$.}
        \label{fig12}
    \end{figure}

The effect of grid stretching on the properties of the Lax-Wendroff method using non-uniform compact scheme is assessed by considering a tangent hyperbolic grid with a higher stretching value of $\gamma=3$. Figures \ref{fig11} and \ref{fig12}, show the property charts of the scheme for the same Peclet numbers, $Pe=0.01$ and $0.02$ for this higher stretched grid case. A careful comparison with the property charts for $\gamma=1.5$ shows that $\alpha_{num}$ improves for the higher grid stretching. Apart from this, one does not observe any significant qualitative differences between the two cases.

The analysis shows that the Lax-Wendroff method is based on non-uniform compact schemes for non-uniform grids can be employed for DNS with much reduced grid requirement than for central difference methods for two reasons. First, the method provides a higher value of $(kh)_{max}$ and hence one would require lesser number of points per wave (PPW). Secondly, the usage of non-uniform grid will allow one to appropriately refine the mesh where needed, coupled with the judicious choice of stretching factor of the mesh. We have already provided details on the resolution requirements and constraints for DNS, LES simulations for the Lax-Wendroff method used along with NUC6. Next, we evaluate the property charts for the case of Lax-Wendroff method that is developed for a non-uniform compact scheme, i.e. applied on a uniform grid as a special case, so that we can compare the two and also check for the consistency of the methods. In this limit, NUC6 scheme reduces to the well-known Lele's sixth order compact scheme \cite{HACM153}.

\section{Application of Lax-Wendroff method for Compact Schemes on Uniform Grids}

Here, we assess the accuracy and consistency of the Lax-Wendroff method that is based on non-uniform compact scheme, for uniform structured grids. In the previous section, it was mentioned that the interior stencil of the sixth order non-uniform compact scheme (NUC6) reduces to sixth order Lele's scheme for uniform grids. Here, we will determine whether the application of Lax-Wendroff strategy yields the same benefits for accuracy, as is noted for the NUC6 scheme.

    \begin{figure}[H]
        \centering
         \includegraphics[scale=0.55]{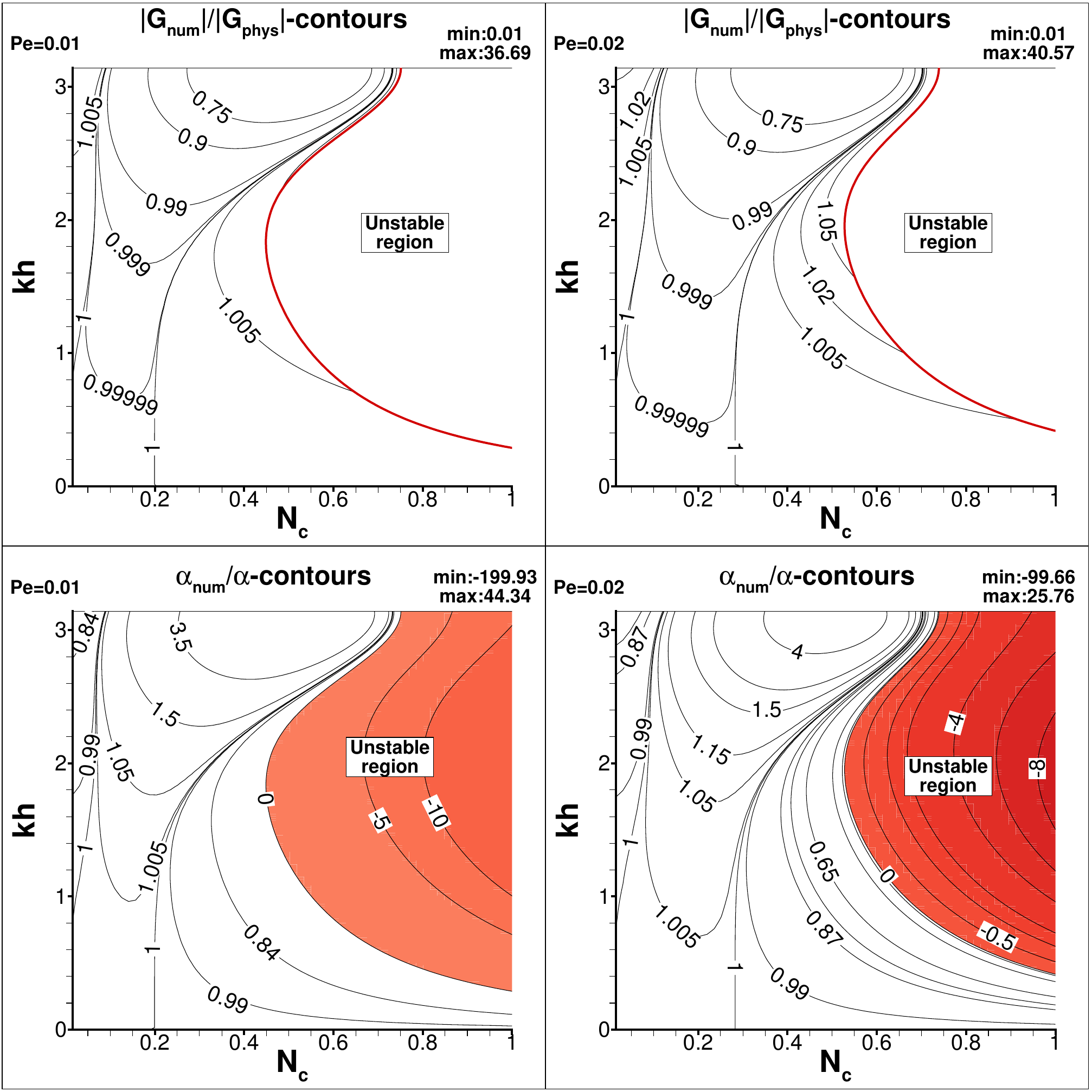}
        \caption{Property charts of the full Lax-Wendroff method of non-uniform compact scheme for the CDE are plotted in $(N_c,kh)$-plane for the Peclet numbers 
				$Pe=0.01$ and $0.02$ when applied for a uniform grid. The ratio of numerical to the corresponding physical amplification factor $\left(\frac{|G_{num}|}{|G_{phys}|}\right)$ and the ratio of numerical to the physical diffusion coefficient $\left(\frac{\alpha_{num}}{\alpha}\right)$ are shown. Regions of numerical instability are as marked in the panels.}
        \label{fig13}
    \end{figure}

    \begin{figure}[H]
        \centering
         \includegraphics[scale=0.55]{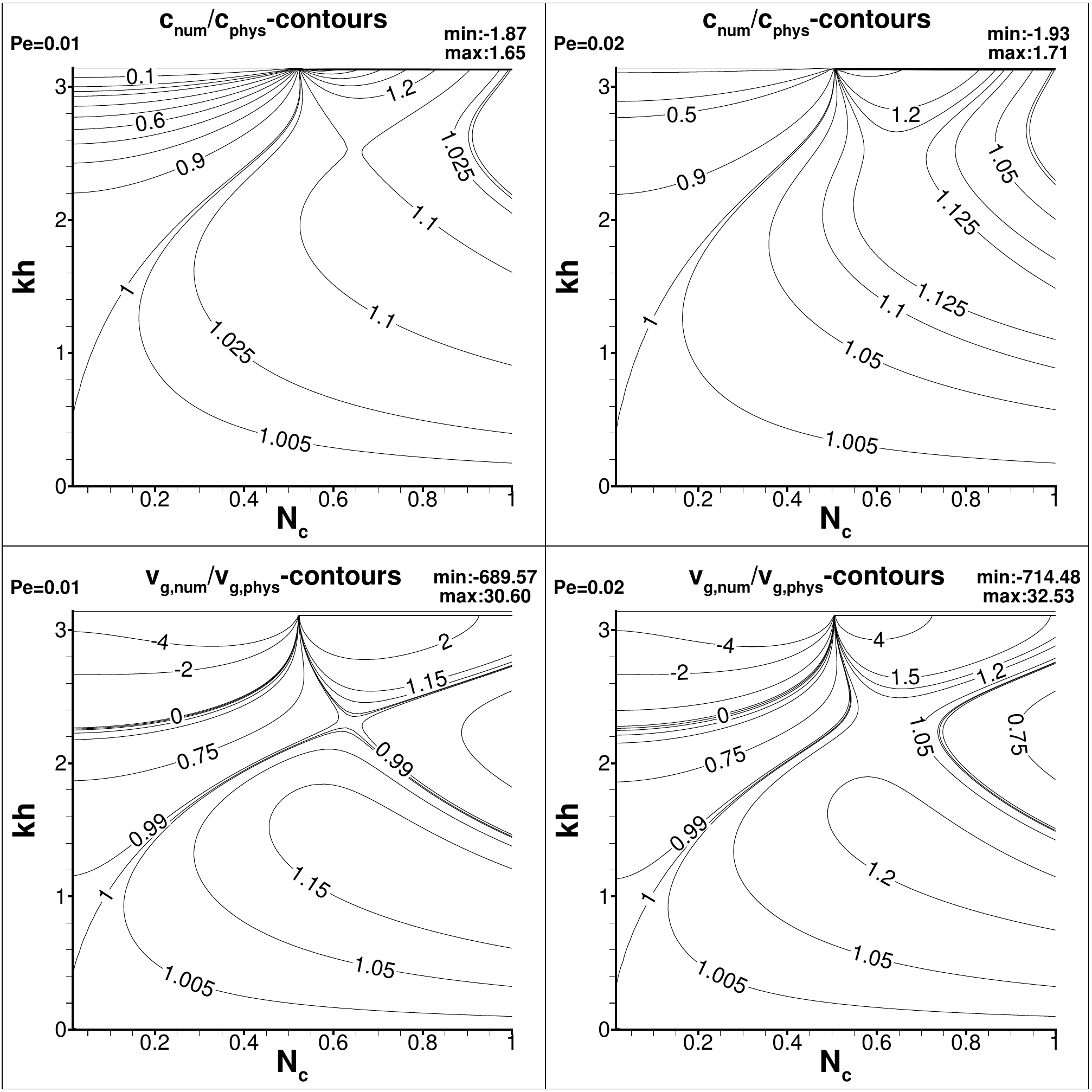}
        \caption{Property charts of the full Lax-Wendroff method of non-uniform compact scheme for the CDE are plotted in $(N_c,kh)$-plane for the Peclet numbers $Pe=0.01$ and $0.02$ when applied for a uniform grid. The ratio of numerical to the corresponding physical phase speed ${c_{num}}/{c_{phys}}$ and the ratio of numerical to the physical group velocity ${V_{g,N}}/{V_{g,phys}}$ are shown.}
        \label{fig14}
    \end{figure}

Figures \ref{fig13} and \ref{fig14} show the property charts for $Pe=0.01$ and $0.02$, respectively, using the Lax-Wendroff method for Lele's sixth order compact scheme \cite{HACM153} for a structured uniform grid. One notes striking similarities in the properties between the present uniform grid and the lesser stretched non-uniform grid results for $\gamma=1.5$ case using NUC6. This demonstrates that benefits of the developed Lax-Wendroff method with the non-uniform compact scheme is retained for the special case of the uniform grid. In Fig. \ref{fig13}, one notices that increasing the Peclet number from $Pe = 0.01$ to $0.02$, increases the admissible $N_c$ value, both with respect to $|G_{num}|/|G_{phys}|$ and $\alpha_{num}/\alpha$. For the uniform grid, one again notes the extended vertical range of the $|G_{num}|/|G_{phys}| = 1$ contour that gives a higher value of $(kh)_{max}$, i.e. higher resolution. In Fig. \ref{fig14}, one notices the dispersion properties to be quite similar to those shown in Fig. \ref{fig12} for the NUC6 compact scheme using the non-uniform tangent hyperbolic grid.  

\section{Quantification of DNS, LES and DRP Metrics for Compact Scheme Based Lax-Wendroff methods}

In this section, we quantify the DRP metrics for the compact scheme based Lax-Wendroff methods for the purpose of performing high accuracy simulations like DNS and LES. For these high accuracy calculations, the normalized properties $\frac{|G_{num}|}{|G_{phys}|}$, $\frac{\alpha_{num}}{\alpha}$, $\frac{c_{num}}{c_{phys}}$ and $\frac{V_{g,{num}}}{V_{g,phys}}$ should be as close to unity as possible, in order to minimize errors due to dissipation, dispersion and signal propagation \cite{HACM259, B118}. In regions where $\frac{|G_{num}|}{|G_{phys}|}$ deviates farther from unity value, errors can appear due to strong/weak numerical diffusion when compared to physical diffusion, even in the stable region. Similarly, errors can also appear due to a lack of satisfaction of dispersion relation, resulting in dispersion and signal propagation errors. These aspects are quantified here for the developed compact scheme based Lax-Wendroff methods for 1D CDE with the help of GSA. 

\begin{figure}[H]
        \centering
         \includegraphics[scale=0.55]{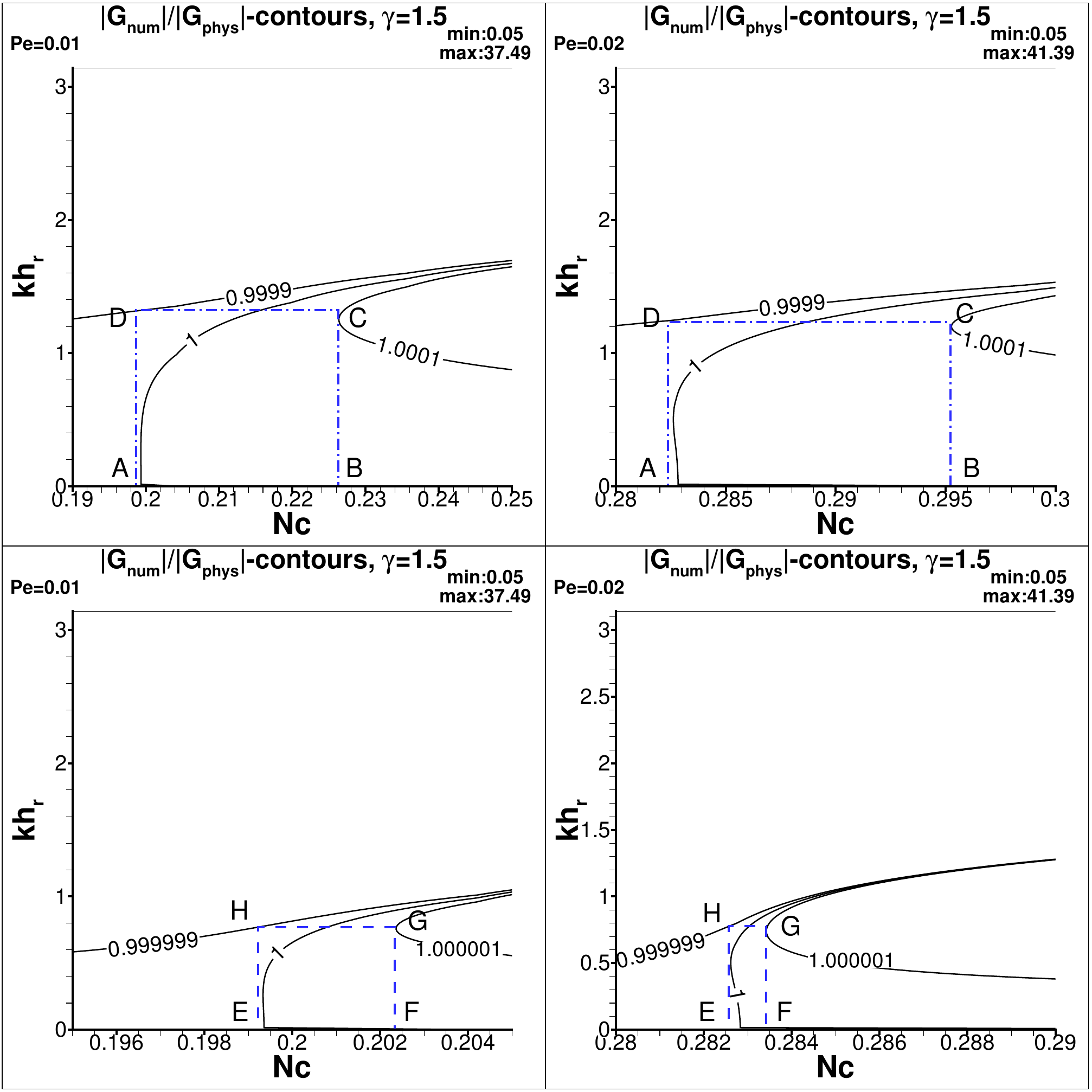}
        \caption{Zones of coarse (top) and refined calculations (bottom) for the non uniform compact scheme based Lax-Wendroff method for all the terms in CDE, for the Peclet numbers $Pe=0.01$, $0.02$ and non-uniform grid with $\gamma=1.5$. The zones are evaluated near the contour with $|G_{num}|/|G_{phys}|=1$. Regions bounded by rectangles ABCD, EFGH denote the coarse and refined calculation zones, for admissible error tolerances as $|1-|G_{num}|/|G_{phys}|| \le 10^{-4}$ and $|1-|G_{num}|/|G_{phys}|| \le 10^{-6}$, respectively.}
				\label{fig15}
    \end{figure}

Figure \ref{fig15} shows a zoomed view of the region around the contour-line $\frac{|G_{num}|}{|G_{phys}|}=1$ intersecting the $N_c$ axis as noted in Fig. \ref{fig9}, for $Pe=0.01$ and $0.02$. This case corresponds to a non-uniform grid with $\gamma=1.5$. In order to quantify the DRP metrics two deviations of $10^{-4}$ and $10^{-6}$ from unity are considered and a region bounding the deviations is determined. In the figure, the top frames correspond to higher deviation (conversely lower resolution) and the bottom frames correspond to the lower deviation case (higher resolution), respectively. One also notes regions marked by (blue) boxes and denoted as ABCD, EFGH in the top and bottom frames, respectively, which identify the acceptable ranges of $N_c$ and $(kh)_{max}$. The boxes are used to quantify the maximum resolution of the scheme and the corresponding deviations in $\alpha_{num}$, $c_{num}$ and $V_{g,num}$, respectively, for the chosen tolerance for error in $\frac{|G_{num}|}{|G_{phys}|}$. We note that the height of the boxes denotes the maximum wavenumber $(kh)_{max}$ that can be resolved by admitting the deviations/tolerance. By comparing the top and bottom panels, one notes that $(kh)_{max}$ decreases as the tolerances are tightened. This implies that a reduction in tolerances necessitates finer grids to resolve the scales up to $(kh)_{max}$. Further, one notes that the $N_c$ range also reduces as the tolerance is reduced. This behavior is expected as the errors due to numerical schemes decrease as one goes towards the continuum limit $kh=0$. To quantify the grid required for obtaining $(kh)_{max}$ resolution, the PPW is determined, vis-$\grave{a}$-vis the Nyquist criterion of $(kh)_{Nyquist} = \pi$ (that requires 3 points for the highest resolved wavenumber \cite{HACM} by the Fourier spectral method). Hence, one requires $(2\pi/ (kh)_{max} + 1)$ points per fully resolved wave. The $(kh)_{max}$ and PPW are provided in Tables \ref{tab_quant3} and \ref{tab_quant4}. One notes $(kh)_{max} = 1.3205$ for $Pe = 0.01$ for the coarse resolution case of $|1-|G_{num}|/|G_{phys}|| \le 10^{-4}$ which reduces to $0.7706$ for the refined calculation. As a result, the PPW value for the coarse calculation is 5, that increases to 9 for the refined calculation. In addition to this data, the maximum variation in $c_{num}/c$ and $V_{g,num}/V_{g,phys}$ are also tabulated, thus, determining the DRP properties.

\begin{figure}[H]
        \centering
         \includegraphics[scale=0.55]{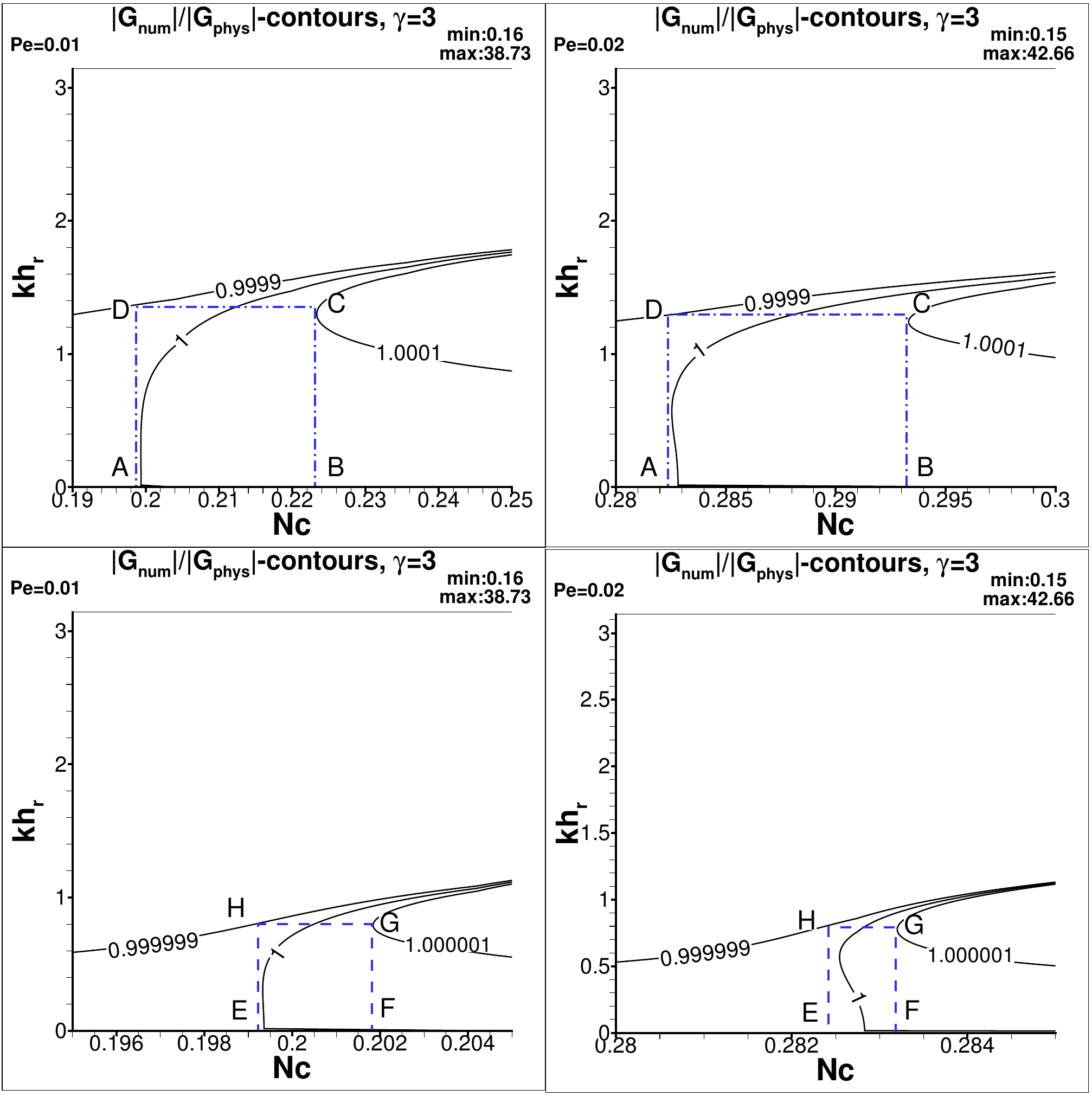}
        \caption{Zones of coarse (top) and refined calculations (bottom) for the non uniform compact scheme based Lax-Wendroff method for all the terms in CDE, for the Peclet numbers $Pe=0.01$, $0.02$ and non-uniform grid with $\gamma=3$. The zones are evaluated near the contour with  $|G_{num}|/|G_{phys}|=1$. Regions bounded by rectangles ABCD, EFGH denote the coarse and refined calculation zones, for admissible error tolerances as $|1-|G_{num}|/|G_{phys}|| \le 10^{-4}$ and $|1-|G_{num}|/|G_{phys}|| \le 10^{-6}$, respectively.}
				\label{fig16}
    \end{figure}

Figure \ref{fig16} shows the zoomed view of the region around the contour-line $\frac{|G_{num}|}{|G_{phys}|}=1$ for non-uniform grid with $\gamma=3$ as noted in Fig. \ref{fig11}. The top frames correspond to higher deviation and the bottom frames correspond to the lower deviation case, respectively. Boxes ABCD and EFGH correspond to the regions in which the tolerances are maintained for the top and bottom frames, respectively. Similar behavior as noted for $\gamma=1.5$ is observed for the tolerances reduced, i.e. $(kh)_{max}$ and admissible $N_c$ ranges reduce. One notes that $(kh)_{max}$ and admissible $N_c$ range values are similar to the coarser non-uniform grid ($\gamma=1.5$). This is attributed to using $h_r$ as the reference length scale for determining $N_c$ and $Pe$, thereby normalizing the properties for different non-uniform grids. The similarity in the properties of the developed non-uniform compact scheme based Lax-Wendroff method for non-uniform grids demonstrate the potential and applicability of the scheme.

Tables \ref{tab_quant3} and \ref{tab_quant4} present the admissible $N_c$ range, $(kh)_{max}$ for the tolerance limits $|1-|G_{num}|/|G_{phys}||\le10^{-4}$ and $|1-|G_{num}|/|G_{phys}||\le10^{-6}$ for the non-uniform compact scheme and Lele's scheme for $Pe=0.01$ and $0.02$, respectively. For the non-uniform compact scheme, the results are determined for the grid with stretching parameter $\gamma=1.5$. It should be noted that the parameter $kh_r$ is applicable for non-uniform grids, whereas for the uniform grid case it is simply $kh$. The resolution limits and the restricted range of $N_c$ values for which the limits are applicable in the tables show a 3 times benefit for the compact schemes over the explicit counterparts. This is once again attributed to the spectral-like resolution of the compact schemes. It is also interesting to note that the resolution limits and $N_c$ range are similar when one considers either non-uniform grids with lower stretching parameter $\gamma$ values or when using uniform grids. This demonstrates the robustness of the developed Lax-Wendroff method using compact differences. It is also interesting to note that only one $N_c$ value is present where $\alpha_{num}=\alpha$ in Figs. \ref{fig9}, \ref{fig11} and \ref{fig13} and one can only use this $N_c$ value. However, based on the result presented here for the central difference case, the higher range of $N_c$ is recommended, because of the better dispersion and signal propagation properties with lower errors given by, $|1-{c_{num}\over c}|$ and $|1-{V_{g,num} \over c}|$.

\begin{table}[h!]
\centering
\resizebox{\columnwidth}{!}{ \begin{tabular}{|c|c|c|c|c|c|c|c|c|c|c|c|c|}
 \hline
Simulation &\multicolumn{2}{c|}{$N_c$} & \multicolumn{2}{c|}{Maximum} & \multicolumn{2}{c|}{Points per} & \multicolumn{2}{c|}{$\alpha_{num}$/ $\alpha$} & \multicolumn{2}{c|}{$c_{num}$/$c$} & \multicolumn{2}{c|}{$V_{g,num}$/$c$} \\
 type & \multicolumn{2}{c|}{} & \multicolumn{2}{c|}{resolution $(kh)_{max}$} & \multicolumn{2}{c|}{wave (PPW)} & \multicolumn{2}{c|}{range} & \multicolumn{2}{c|}{range} & \multicolumn{2}{c|}{range} \\
\cline{2-13}& NUC6 & Lele & NUC6 & Lele & NUC6 & Lele & NUC6 & Lele & NUC6 & Lele & NUC6 & Lele \\ [0.5ex]
 \hline
 Coarse & 0.2126 & 0.2132 & 1.3205 & 1.292 &5 &5& 0.9947-1.006 & 0.9945-1.006 & 1-1.0122 & 1-1.012& 1-1.0241 &1-1.0247\\
 Refined& 0.2008 & 0.2009 & 0.7706 & 0.7543 &9&9& 0.9998-1.0002 & 0.9998-1.0002& 1-1.0041 &1-1.0038& 1-1.0118&1-1.0111\\[1ex]
 \hline
 \end{tabular}}
 \caption{Comparison of DRP properties between Lax-Wendroff NUC6 ($\gamma=1.5$) and Lax-Wendroff Lele schemes for coarse ($|1-|G_{num}|/|G_{phys}||\le10^{-4}$) and refined simulations ($|1-|G_{num}|/|G_{phys}||\le10^{-6}$) for the 1D CDE using $Pe=0.01$. }
 \label{tab_quant3}
\end{table}

\begin{table}[h!]
\centering
\resizebox{\columnwidth}{!}{ \begin{tabular}{|c|c|c|c|c|c|c|c|c|c|c|c|c|}
 \hline
Simulation &\multicolumn{2}{c|}{$N_c$} & \multicolumn{2}{c|}{Maximum} & \multicolumn{2}{c|}{Points per} & \multicolumn{2}{c|}{$\alpha_{num}$/ $\alpha$} & \multicolumn{2}{c|}{$c_{num}$/$c$} & \multicolumn{2}{c|}{$V_{g,num}$/$c$} \\
 type & \multicolumn{2}{c|}{} & \multicolumn{2}{c|}{resolution $(kh)_{max}$} & \multicolumn{2}{c|}{wave (PPW)} & \multicolumn{2}{c|}{range} & \multicolumn{2}{c|}{range} & \multicolumn{2}{c|}{range} \\
\cline{2-13}& NUC6 & Lele & NUC6 & Lele & NUC6 & Lele & NUC6 & Lele & NUC6 & Lele & NUC6 & Lele \\ [0.5ex]
 \hline
 Coarse & 0.2888 & 0.2888 & 1.2323 & 1.2204 &6&6& 0.9968-1.003 & 0.9973-1.0034 & 1-1.0208 & 1-1.0205& 1-1.056 &1-1.0548\\
 Refined& 0.283 & 0.283 & 0.7764 & 0.7509 &9&9& 0.9999-1.00009 & 0.9999-1.00009& 1-1.008 &  1-1.0075& 1-1.0241&1-1.0223\\[1ex]
 \hline
 \end{tabular}}
 \caption{Comparison of DRP properties between Lax-Wendroff NUC6 ($\gamma=1.5$) and Lax-Wendroff Lele schemes for coarse ($|1-|G_{num}|/|G_{phys}||\le10^{-4}$) and refined simulations ($|1-|G_{num}|/|G_{phys}|| \le 10^{-6}$) for the 1D CDE using $Pe=0.02$.}
 \label{tab_quant4}
\end{table}

\begin{table}[h!]
\centering
\resizebox{\columnwidth}{!}{ \begin{tabular}{|c|c|c|c|c|c|c|c|c|c|c|c|c|}
 \hline
Simulation &\multicolumn{2}{c|}{$N_c$} & \multicolumn{2}{c|}{Maximum} & \multicolumn{2}{c|}{Points per} & \multicolumn{2}{c|}{$\alpha_{num}$/ $\alpha$} & \multicolumn{2}{c|}{$c_{num}$/$c$} & \multicolumn{2}{c|}{$V_{g,num}$/$c$} \\
 type & \multicolumn{2}{c|}{} & \multicolumn{2}{c|}{resolution $(kh)_{max}$} & \multicolumn{2}{c|}{wave (PPW)} & \multicolumn{2}{c|}{range} & \multicolumn{2}{c|}{range} & \multicolumn{2}{c|}{range} \\
\cline{2-13}& NUC6 & Lele & NUC6 & Lele & NUC6 & Lele & NUC6 & Lele & NUC6 & Lele & NUC6 & Lele \\ [0.5ex]
 \hline
 Coarse & 0.2109 & 0.2132 & 1.3519 & 1.292  &5&5& 0.9949-1.005  & 0.9945-1.006 & 1-1.0124 &1-1.012 & 1-1.0235 &1-1.0247\\
 Refined& 0.2004 & 0.2009 & 0.8021 & 0.7543 &8&9& 0.9998-1.0002 & 0.9998-1.0002& 1-1.0042 &1-1.0038& 1-1.0124 &1-1.0111\\[1ex]
 \hline
 \end{tabular}}
 \caption{Comparison of DRP properties between Lax-Wendroff NUC6 ($\gamma=3$) and Lax-Wendroff Lele schemes for coarse ($|1-|G_{num}|/|G_{phys}||\le10^{-4}$) and refined simulations ($|1-|G_{num}|/|G_{phys}||\le10^{-6}$) for the 1D CDE using $Pe=0.01$. }
 \label{tab_quant5}
\end{table}

\begin{table}[h!]
\centering
\resizebox{\columnwidth}{!}{ \begin{tabular}{|c|c|c|c|c|c|c|c|c|c|c|c|c|}
 \hline
Simulation &\multicolumn{2}{c|}{$N_c$} & \multicolumn{2}{c|}{Maximum} & \multicolumn{2}{c|}{Points per} & \multicolumn{2}{c|}{$\alpha_{num}$/ $\alpha$} & \multicolumn{2}{c|}{$c_{num}$/$c$} & \multicolumn{2}{c|}{$V_{g,num}$/$c$} \\
 type & \multicolumn{2}{c|}{} & \multicolumn{2}{c|}{resolution $(kh)_{max}$} & \multicolumn{2}{c|}{wave (PPW)} & \multicolumn{2}{c|}{range} & \multicolumn{2}{c|}{range} & \multicolumn{2}{c|}{range} \\
\cline{2-13}& NUC6 & Lele & NUC6 & Lele & NUC6 & Lele & NUC6 & Lele & NUC6 & Lele & NUC6 & Lele \\ [0.5ex]
 \hline
 Coarse & 0.2878& 0.2888 & 1.2952 & 1.2204 &5&6& 0.9972-1.003   & 0.9973-1.0034 & 1-1.0225 & 1-1.0205& 1-1.0594 &1-1.0548\\
 Refined& 0.2828& 0.283  & 0.7922 & 0.7509 &8&9& 0.99992-1.00007& 0.9999-1.00009& 1-1.0085 & 1-1.0075& 1-1.0252 &1-1.0223\\[1ex]
 \hline
 \end{tabular}}
 \caption{Comparison of DRP properties between Lax-Wendroff NUC6 ($\gamma=3$) and Lax-Wendroff Lele schemes for coarse ($|1-|G_{num}|/|G_{phys}||\le10^{-4}$) and refined simulations ($|1-|G_{num}|/|G_{phys}|| \le 10^{-6}$) for the 1D CDE using $Pe=0.02$.}
 \label{tab_quant6}
\end{table}

In Tables \ref{tab_quant5} and \ref{tab_quant6}, the admissible $N_c$ range and $(kh)_{max}$ are presented primarily for the tolerance limits $|1-|G_{num}|/|G_{phys}||\le10^{-4}$ and $|1-|G_{num}|/|G_{phys}||\le10^{-6}$ to compare the performance characteristics of the non-uniform compact scheme (NUC6 scheme with $\gamma = 3$) and Lele's scheme for $Pe=0.01$ and $0.02$, respectively. For the non-uniform compact scheme, the results are determined for the grid with stretching parameter $\gamma=3$, which has been shown earlier to be better with the stretching parameter increasing. The $x$-axis has been indicated by the parameter $kh_r$ for the non-uniform grid, which for the uniform grid case is simply $kh$. Compared to the lower stretching parameter case of $\gamma = 1.5$, for this higher stretched grid, the value of $N_c$ remains more or less the same for the NUC6 scheme. However, the benefit in PPW reduction is noted for $\gamma =3$, as compared to the $\gamma = 1.5$ case, as $(kh)_{max}$ increases with grid stretching. Increasing the Peclet number to 0.02, mildly improves the performance with respect to $(kh)_{max}$. With such improvements, one clearly notes the superior parameters of the NUC6 scheme, over Lele's scheme as can be seen in these tables. 

\section{Summary and Conclusions}
The GSA of numerical methods for the compact schemes reported here, demonstrates that the DRP property shows better performance features for DNS/LES for the Lax-Wendroff method. One of the major advantages noted for the explicit central scheme is the existence of two distinct values of $N_c$ available for which DNS and LES can be performed. It was noted that the second value of $N_c$ is not only significantly higher but also the other properties given in terms of $(kh)_{max}$, points per wave (PPW) required to perform DNS and LES are significantly lower. In contrast, the admissible $N_c$ value for the compact schemes reported in this part for uniform and non-uniform grids is comparatively lower, implying a necessary smaller time step for the compact scheme. In compensating this disadvantage for the compact schemes, one notices significant better and higher value of $(kh)_{max}$ for the compact schemes over the explicit second order scheme. This will automatically ensure a significantly large reduction in number of points for the compact scheme. Additionally, one also notices significantly improved DRP properties. 

In this paper, we have drawn comparisons between the numerical performance of the NUC6 scheme and Lele's sixth order scheme that one would obtain as a limiting case of NUC6 scheme applied on a uniform grid. It has been demonstrated via the property charts obtained by GSA that the NUC6 scheme outperforms Lele's scheme in terms of the maximum admissible $N_c$ value and the higher wavenumber values to which scales are resolved, $(kh)_{max}$. This fact is more pronounced on increasing the grid stretching parameter from $\gamma = 1.5$ to $3$. The dispersion properties are found to be almost identical for NUC6 scheme compared with Lele's compact scheme. Upon quantifying the resolution required for DNS and LES, it is noted that the resolution requirements of Lele's scheme are higher than NUC6 scheme i.e. the number of points required to resolve a wave, (PPW) is two times higher in Lele's compact scheme compared to NUC6 scheme. Here also, the improved $(kh)_{max}$ and maximum admissible $N_c$ are noted for both stretching parameters in the case of NUC6 scheme compared to Lele's scheme.

\begin{figure}[H]
        \centering
         \includegraphics[scale=1.00]{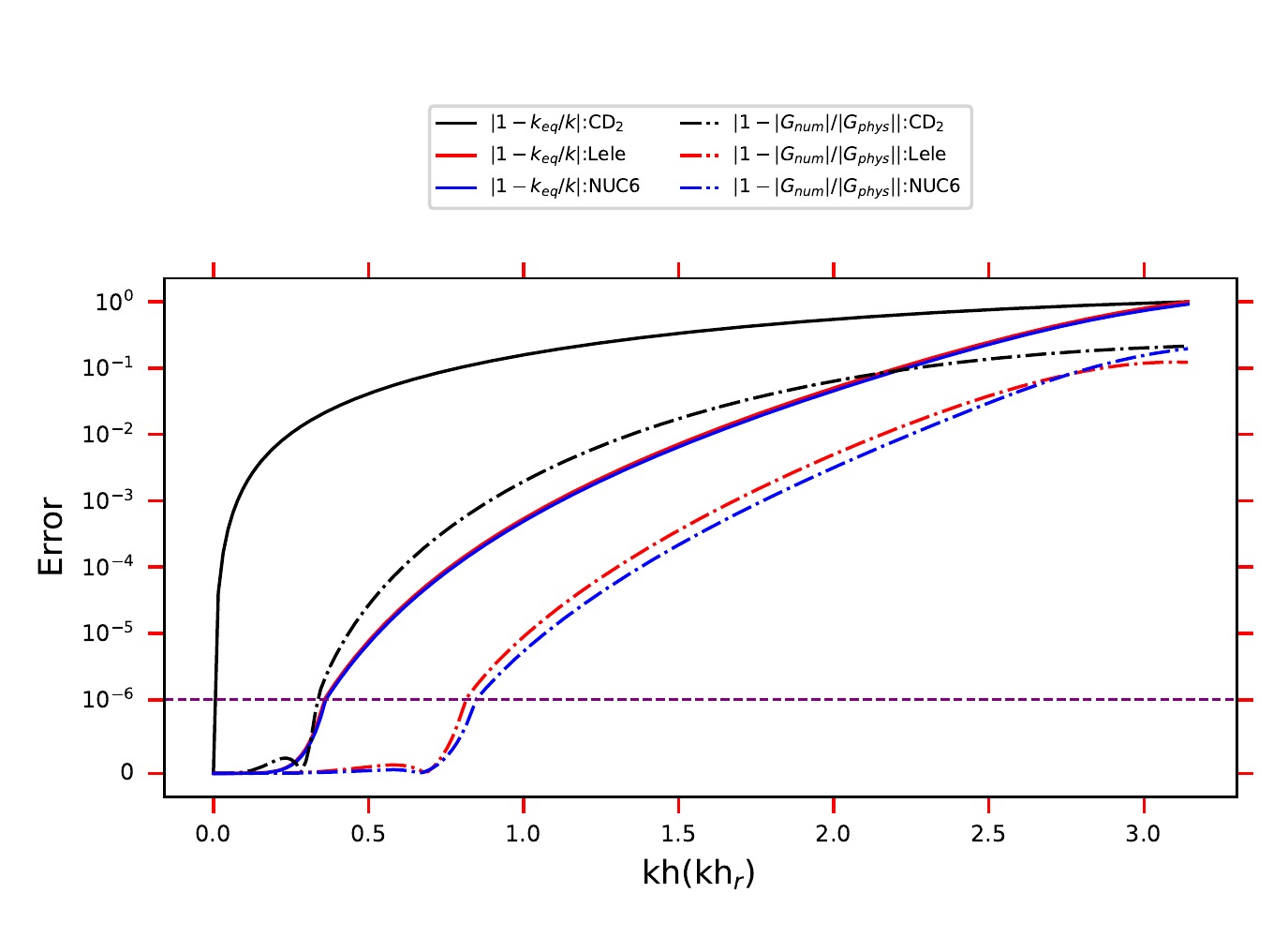}
        \caption{Comparison of explicit and compact schemes used for Lax-Wendroff method, in terms of the traditional measure of resolution via the discretization of first spatial derivative ($k_{eq}/k$) and the proposed quantification here in terms of normalized numerical amplification factor ($|G_{num}|/|G_{phys}|$). To make a more refined distinctions among the methods, note the choice of the ordinate and its depiction in the logarithmic scale. It is readily noted that the superiority of the present quantification, as opposed to the tradition metric used in the literature with $k_{eq}/k$.}
				\label{fig17}
    \end{figure}

The above summary is pictorially depicted in Fig. \ref{fig17}. In this, we have plotted the traditional measure of spatial resolution, $k_{eq}/k$ and the normalized numerical amplification factor, $|G_{num}|/|G_{phys}|$ against $kh_r$ for the non-uniform scheme and $kh$ for Lele's scheme and the central difference method. It is evident that the compact schemes (Lele and NUC6) are performing much better than the explicit central difference method from the lower values of error noted in both quantities. Comparison between the non-uniform compact scheme and its uniform equivalent, reiterates that NUC6 performs better than Lele's scheme, an observation that is more pronounced for higher values of grid stretching parameter, $\gamma$.

Thus, we have investigated the applicability of the newly developed combinations of Lax-Wendroff method with Lele's sixth order compact scheme and NUC6 scheme using GSA. The Lax-Wendroff method which has second-order accuracy in time without any spurious modes has interesting features for second order central differencing scheme, as well as for implicit compact schemes on uniform and non-uniform grids. While the grid point requirements for the explicit scheme are higher, it allows to select larger time steps comparatively. In contrast, the compact schemes take relatively smaller time steps but allow resolving scales for DNS and LES with fewer points per wave. 

\newpage
\section*{Appendix-1: Sixth Order Non-Uniform Grid Based Compact Scheme for Higher Derivatives}
The coefficients of sixth order non-uniform grid based compact scheme for second derivative is obtained from Eq. \eqref{der2}.
The Taylor series expansion is used in Eq. \eqref{der2} and coefficients of various order are matched to obtain a linear relation between $p^2_{-1}$, $p^2_{1}$, $s^2_{-2}$, $s^2_{-1}$, $s^2_{0}$, $s^2_{1}$ and $s^2_{2}$, which is given as

\begin{equation}
    s^2_{-2} + s^2_{-1} + s^2_0 + s^2_{1} + s^2_{2} = 0,
\end{equation}
and
\begin{equation}
    \left[\begin{array}{cccccc}
        -\mathrm{h_{ll}} & -\mathrm{h_{l}} & \mathrm{h_{r}} & \mathrm{h_{rr}} & 0 & 0\\
        \frac{{\mathrm{h_{ll}}}^2}{2} & \frac{{\mathrm{h_{l}}}^2}{2} & \frac{{\mathrm{h_{r}}}^2}{2} & \frac{{\mathrm{hrr}}^2}{2} & -1 & -1\\
        -\frac{{\mathrm{h_{ll}}}^3}{6} & -\frac{{\mathrm{h_{l}}}^3}{6} & \frac{{\mathrm{h_{r}}}^3}{6} & \frac{{\mathrm{hrr}}^3}{6} & \mathrm{h_{l}} & -\mathrm{h_{r}}\\
        \frac{{\mathrm{h_{ll}}}^4}{24} & \frac{{\mathrm{h_{l}}}^4}{24} & \frac{{\mathrm{h_{r}}}^4}{24} & \frac{{\mathrm{hrr}}^4}{24} & -\frac{{\mathrm{h_{l}}}^2}{2} & -\frac{{\mathrm{h_{r}}}^2}{2}\\
        -\frac{{\mathrm{h_{ll}}}^5}{120} & -\frac{{\mathrm{h_{l}}}^5}{120} & \frac{{\mathrm{h_{r}}}^5}{120} & \frac{{\mathrm{hrr}}^5}{120} & \frac{{\mathrm{h_{l}}}^3}{6} & -\frac{{\mathrm{h_{r}}}^3}{6}\\
        \frac{{\mathrm{h_{ll}}}^6}{720} & \frac{{\mathrm{h_{l}}}^6}{720} & \frac{{\mathrm{h_{r}}}^6}{720} & \frac{{\mathrm{hrr}}^6}{720} & -\frac{{\mathrm{h_{l}}}^4}{24} & -\frac{{\mathrm{h_{r}}}^4}{24}
    \end{array}\right]
    \left[\begin{array}{c} s^2_{-2}\\ s^2_{-1}\\ s^2_{1}\\ s^2_{2}\\ p^2_{-1}\\ p^2_{1} \end{array}\right] =
    \left[\begin{array}{c} 0     \\ 1     \\ 0    \\ 0    \\ 0     \\ 0      \end{array}\right]. \label{Eq:Ax_b2}
\end{equation}

\noindent The linear system of equations is solved using MATLAB\textsuperscript{\textregistered} Symbolic Math Toolbox\textsuperscript{\texttrademark} and the solution of this linear system of equations is provided as,
\\
\\
\noindent $p^2_{-1} =(3h_l^2h_{ll}^2h_r^3 + 3h_l^2h_{ll}^2h_r^2h_{rr} - 3h_l^2h_{ll}^2h_rh_{rr}^2 + 4h_l^2h_{ll}h_r^4 + h_l^2h_{ll}h_r^3h_{rr} - 3h_l^2h_{ll}h_r^2h_{rr}^2 - 4h_l^2h_r^4h_{rr} + 3h_l^2h_r^3h_{rr}^2 + 4h_lh_{ll}^2h_r^4 + h_lh_{ll}^2h_r^3h_{rr} - 3h_lh_{ll}^2h_r^2h_{rr}^2 + 5h_lh_{ll}h_r^5 - 3h_lh_{ll}h_r^4h_{rr} - h_lh_{ll}h_r^3h_{rr}^2 - 5h_lh_r^5h_{rr} + 4h_lh_r^4h_{rr}^2 - 4h_{ll}^2h_r^4h_{rr} + 3h_{ll}^2h_r^3h_{rr}^2 - 5h_{ll}h_r^5h_{rr} + 4h_{ll}h_r^4h_{rr}^2)/((h_l + h_r)(- 10h_l^4h_{ll}h_r + 5h_l^4h_{ll}h_{rr} - 15h_l^4h_r^2 + 10h_l^4h_rh_{rr} + 8h_l^3h_{ll}^2h_r - 4h_l^3h_{ll}^2h_{rr} - 13h_l^3h_{ll}h_r^2 + h_l^3h_{ll}h_rh_{rr} + 4h_l^3h_{ll}h_{rr}^2 - 33h_l^3h_r^3 + 13h_l^3h_r^2h_{rr} + 8h_l^3h_rh_{rr}^2 + 19h_l^2h_{ll}^2h_r^2 - 7h_l^2h_{ll}^2h_rh_{rr} - 3h_l^2h_{ll}^2h_{rr}^2 + 13h_l^2h_{ll}h_r^3 - 20h_l^2h_{ll}h_r^2h_{rr} + 7h_l^2h_{ll}h_rh_{rr}^2 - 15h_l^2h_r^4 - 13h_l^2h_r^3h_{rr} + 19h_l^2h_r^2h_{rr}^2 + 8h_lh_{ll}^2h_r^3 + 7h_lh_{ll}^2h_r^2h_{rr} - 9h_lh_{ll}^2h_rh_{rr}^2 + 10h_lh_{ll}h_r^4 + h_lh_{ll}h_r^3h_{rr} - 7h_lh_{ll}h_r^2h_{rr}^2 - 10h_lh_r^4h_{rr} + 8h_lh_r^3h_{rr}^2 + 4h_{ll}^2h_r^3h_{rr} - 3h_{ll}^2h_r^2h_{rr}^2 + 5h_{ll}h_r^4h_{rr} - 4h_{ll}h_r^3h_{rr}^2))$,
\\
\\
\noindent $p^2_{1} = (- 5h_l^5h_{ll}h_r - 5h_l^5h_{ll}h_{rr} + 5h_l^5h_rh_{rr} + 4h_l^4h_{ll}^2h_r + 4h_l^4h_{ll}^2h_{rr} - 4h_l^4h_{ll}h_r^2 - 3h_l^4h_{ll}h_rh_{rr} - 4h_l^4h_{ll}h_{rr}^2 + 4h_l^4h_r^2h_{rr} + 4h_l^4h_rh_{rr}^2 + 3h_l^3h_{ll}^2h_r^2 - h_l^3h_{ll}^2h_rh_{rr} + 3h_l^3h_{ll}^2h_{rr}^2 + h_l^3h_{ll}h_r^2h_{rr} + h_l^3h_{ll}h_rh_{rr}^2 + 3h_l^3h_r^2h_{rr}^2 - 3h_l^2h_{ll}^2h_r^2h_{rr} - 3h_l^2h_{ll}^2h_rh_{rr}^2 + 3h_l^2h_{ll}h_r^2h_{rr}^2 - 3h_lh_{ll}^2h_r^2h_{rr}^2)/((h_l + h_r)(- 10h_l^4h_{ll}h_r + 5h_l^4h_{ll}h_{rr} - 15h_l^4h_r^2 + 10h_l^4h_rh_{rr} + 8h_l^3h_{ll}^2h_r - 4h_l^3h_{ll}^2h_{rr} - 13h_l^3h_{ll}h_r^2 + h_l^3h_{ll}h_rh_{rr} + 4h_l^3h_{ll}h_{rr}^2 - 33h_l^3h_r^3 + 13h_l^3h_r^2h_{rr} + 8h_l^3h_rh_{rr}^2 + 19h_l^2h_{ll}^2h_r^2 - 7h_l^2h_{ll}^2h_rh_{rr} - 3h_l^2h_{ll}^2h_{rr}^2 + 13h_l^2h_{ll}h_r^3 - 20h_l^2h_{ll}h_r^2h_{rr} + 7h_l^2h_{ll}h_rh_{rr}^2 - 15h_l^2h_r^4 - 13h_l^2h_r^3h_{rr} + 19h_l^2h_r^2h_{rr}^2 + 8h_lh_{ll}^2h_r^3 + 7h_lh_{ll}^2h_r^2h_{rr} - 9h_lh_{ll}^2h_rh_{rr}^2 + 10h_lh_{ll}h_r^4 + h_lh_{ll}h_r^3h_{rr} - 7h_lh_{ll}h_r^2h_{rr}^2 - 10h_lh_r^4h_{rr} + 8h_lh_r^3h_{rr}^2 + 4h_{ll}^2h_r^3h_{rr} - 3h_{ll}^2h_r^2h_{rr}^2 + 5h_{ll}h_r^4h_{rr} - 4h_{ll}h_r^3h_{rr}^2))$,
\\
\\
\noindent $s^2_{-2} = -(6h_lh_r(5h_l^4h_r^2 + 5h_l^4h_rh_{rr} - 5h_l^4h_{rr}^2 + 11h_l^3h_r^3 + 6h_l^3h_r^2h_{rr} - 4h_l^3h_rh_{rr}^2 - 4h_l^3h_{rr}^3 + 5h_l^2h_r^4 - 6h_l^2h_r^3h_{rr} + 9h_l^2h_r^2h_{rr}^2 - 6h_l^2h_rh_{rr}^3 - 5h_lh_r^4h_{rr} - 4h_lh_r^3h_{rr}^2 + 6h_lh_r^2h_{rr}^3 - 5h_r^4h_{rr}^2 + 4h_r^3h_{rr}^3))/(h_{ll}(h_{ll} + h_r)(h_{ll} + h_{rr})(h_l - h_{ll})(- 10h_l^4h_{ll}h_r + 5h_l^4h_{ll}h_{rr} - 15h_l^4h_r^2 + 10h_l^4h_rh_{rr} + 8h_l^3h_{ll}^2h_r - 4h_l^3h_{ll}^2h_{rr} - 13h_l^3h_{ll}h_r^2 + h_l^3h_{ll}h_rh_{rr} + 4h_l^3h_{ll}h_{rr}^2 - 33h_l^3h_r^3 + 13h_l^3h_r^2h_{rr} + 8h_l^3h_rh_{rr}^2 + 19h_l^2h_{ll}^2h_r^2 - 7h_l^2h_{ll}^2h_rh_{rr} - 3h_l^2h_{ll}^2h_{rr}^2 + 13h_l^2h_{ll}h_r^3 - 20h_l^2h_{ll}h_r^2h_{rr} + 7h_l^2h_{ll}h_rh_{rr}^2 - 15h_l^2h_r^4 - 13h_l^2h_r^3h_{rr} + 19h_l^2h_r^2h_{rr}^2 + 8h_lh_{ll}^2h_r^3 + 7h_lh_{ll}^2h_r^2h_{rr} - 9h_lh_{ll}^2h_rh_{rr}^2 + 10h_lh_{ll}h_r^4 + h_lh_{ll}h_r^3h_{rr} - 7h_lh_{ll}h_r^2h_{rr}^2 - 10h_lh_r^4h_{rr} + 8h_lh_r^3h_{rr}^2 + 4h_{ll}^2h_r^3h_{rr} - 3h_{ll}^2h_r^2h_{rr}^2 + 5h_{ll}h_r^4h_{rr} - 4h_{ll}h_r^3h_{rr}^2))$,
\\
\\
\noindent $s^2_{-1} = -(6(- 15h_l^2h_{ll}^2h_r^3 - 15h_l^2h_{ll}^2h_r^2h_{rr} + 15h_l^2h_{ll}^2h_rh_{rr}^2 - 20h_l^2h_{ll}h_r^4 - 5h_l^2h_{ll}h_r^3h_{rr} + 15h_l^2h_{ll}h_r^2h_{rr}^2 + 20h_l^2h_r^4h_{rr} - 15h_l^2h_r^3h_{rr}^2 + 10h_lh_{ll}^3h_r^3 + 10h_lh_{ll}^3h_r^2h_{rr} - 10h_lh_{ll}^3h_rh_{rr}^2 + 5h_lh_{ll}^2h_r^4 - 5h_lh_{ll}^2h_r^3h_{rr} - 15h_lh_{ll}^2h_r^2h_{rr}^2 + 10h_lh_{ll}^2h_rh_{rr}^3 - 10h_lh_{ll}h_r^5 - 15h_lh_{ll}h_r^4h_{rr} + 5h_lh_{ll}h_r^3h_{rr}^2 + 10h_lh_{ll}h_r^2h_{rr}^3 + 10h_lh_r^5h_{rr} + 5h_lh_r^4h_{rr}^2 - 10h_lh_r^3h_{rr}^3 + 4h_{ll}^3h_r^4 + 4h_{ll}^3h_r^3h_{rr} + 4h_{ll}^3h_r^2h_{rr}^2 - 6h_{ll}^3h_rh_{rr}^3 + 5h_{ll}^2h_r^5 + h_{ll}^2h_r^4h_{rr} + h_{ll}^2h_r^3h_{rr}^2 - 4h_{ll}^2h_r^2h_{rr}^3 - 5h_{ll}h_r^5h_{rr} - h_{ll}h_r^4h_{rr}^2 + 4h_{ll}h_r^3h_{rr}^3 + 5h_r^5h_{rr}^2 - 4h_r^4h_{rr}^3))/((h_l + h_r)(h_l + h_{rr})(h_l - h_{ll})(- 10h_l^4h_{ll}h_r + 5h_l^4h_{ll}h_{rr} - 15h_l^4h_r^2 + 10h_l^4h_rh_{rr} + 8h_l^3h_{ll}^2h_r - 4h_l^3h_{ll}^2h_{rr} - 13h_l^3h_{ll}h_r^2 + h_l^3h_{ll}h_rh_{rr} + 4h_l^3h_{ll}h_{rr}^2 - 33h_l^3h_r^3 + 13h_l^3h_r^2h_{rr} + 8h_l^3h_rh_{rr}^2 + 19h_l^2h_{ll}^2h_r^2 - 7h_l^2h_{ll}^2h_rh_{rr} - 3h_l^2h_{ll}^2h_{rr}^2 + 13h_l^2h_{ll}h_r^3 - 20h_l^2h_{ll}h_r^2h_{rr} + 7h_l^2h_{ll}h_rh_{rr}^2 - 15h_l^2h_r^4 - 13h_l^2h_r^3h_{rr} + 19h_l^2h_r^2h_{rr}^2 + 8h_lh_{ll}^2h_r^3 + 7h_lh_{ll}^2h_r^2h_{rr} - 9h_lh_{ll}^2h_rh_{rr}^2 + 10h_lh_{ll}h_r^4 + h_lh_{ll}h_r^3h_{rr} - 7h_lh_{ll}h_r^2h_{rr}^2 - 10h_lh_r^4h_{rr} + 8h_lh_r^3h_{rr}^2 + 4h_{ll}^2h_r^3h_{rr} - 3h_{ll}^2h_r^2h_{rr}^2 + 5h_{ll}h_r^4h_{rr} - 4h_{ll}h_r^3h_{rr}^2)) $,
\\
\\
\noindent $s^2_{1} = -(6(5h_l^5h_{ll}^2 + 10h_l^5h_{ll}h_r - 5h_l^5h_{ll}h_{rr} - 10h_l^5h_rh_{rr} + 5h_l^5h_{rr}^2 - 4h_l^4h_{ll}^3 + 5h_l^4h_{ll}^2h_r - h_l^4h_{ll}^2h_{rr} + 20h_l^4h_{ll}h_r^2 - 15h_l^4h_{ll}h_rh_{rr} + h_l^4h_{ll}h_{rr}^2 - 20h_l^4h_r^2h_{rr} + 5h_l^4h_rh_{rr}^2 + 4h_l^4h_{rr}^3 - 10h_l^3h_{ll}^3h_r + 4h_l^3h_{ll}^3h_{rr} - 15h_l^3h_{ll}^2h_r^2 + 5h_l^3h_{ll}^2h_rh_{rr} + h_l^3h_{ll}^2h_{rr}^2 - 5h_l^3h_{ll}h_r^2h_{rr} - 5h_l^3h_{ll}h_rh_{rr}^2 + 4h_l^3h_{ll}h_{rr}^3 - 15h_l^3h_r^2h_{rr}^2 + 10h_l^3h_rh_{rr}^3 + 10h_l^2h_{ll}^3h_rh_{rr} - 4h_l^2h_{ll}^3h_{rr}^2 + 15h_l^2h_{ll}^2h_r^2h_{rr} - 15h_l^2h_{ll}^2h_rh_{rr}^2 + 4h_l^2h_{ll}^2h_{rr}^3 - 15h_l^2h_{ll}h_r^2h_{rr}^2 + 10h_l^2h_{ll}h_rh_{rr}^3 + 10h_lh_{ll}^3h_rh_{rr}^2 - 6h_lh_{ll}^3h_{rr}^3 + 15h_lh_{ll}^2h_r^2h_{rr}^2 - 10h_lh_{ll}^2h_rh_{rr}^3))/((h_l + h_r)(h_{ll} + h_r)(h_r - h_{rr})(- 10h_l^4h_{ll}h_r + 5h_l^4h_{ll}h_{rr} - 15h_l^4h_r^2 + 10h_l^4h_rh_{rr} + 8h_l^3h_{ll}^2h_r - 4h_l^3h_{ll}^2h_{rr} - 13h_l^3h_{ll}h_r^2 + h_l^3h_{ll}h_rh_{rr} + 4h_l^3h_{ll}h_{rr}^2 - 33h_l^3h_r^3 + 13h_l^3h_r^2h_{rr} + 8h_l^3h_rh_{rr}^2 + 19h_l^2h_{ll}^2h_r^2 - 7h_l^2h_{ll}^2h_rh_{rr} - 3h_l^2h_{ll}^2h_{rr}^2 + 13h_l^2h_{ll}h_r^3 - 20h_l^2h_{ll}h_r^2h_{rr} + 7h_l^2h_{ll}h_rh_{rr}^2 - 15h_l^2h_r^4 - 13h_l^2h_r^3h_{rr} + 19h_l^2h_r^2h_{rr}^2 + 8h_lh_{ll}^2h_r^3 + 7h_lh_{ll}^2h_r^2h_{rr} - 9h_lh_{ll}^2h_rh_{rr}^2 + 10h_lh_{ll}h_r^4 + h_lh_{ll}h_r^3h_{rr} - 7h_lh_{ll}h_r^2h_{rr}^2 - 10h_lh_r^4h_{rr} + 8h_lh_r^3h_{rr}^2 + 4h_{ll}^2h_r^3h_{rr} - 3h_{ll}^2h_r^2h_{rr}^2 + 5h_{ll}h_r^4h_{rr} - 4h_{ll}h_r^3h_{rr}^2))$,
\\
\\
\noindent $s^2_{2} = -(6h_lh_r(- 5h_l^4h_{ll}^2 - 5h_l^4h_{ll}h_r + 5h_l^4h_r^2 + 4h_l^3h_{ll}^3 - 4h_l^3h_{ll}^2h_r - 6h_l^3h_{ll}h_r^2 + 11h_l^3h_r^3 + 6h_l^2h_{ll}^3h_r + 9h_l^2h_{ll}^2h_r^2 + 6h_l^2h_{ll}h_r^3 + 5h_l^2h_r^4 - 6h_lh_{ll}^3h_r^2 - 4h_lh_{ll}^2h_r^3 + 5h_lh_{ll}h_r^4 - 4h_{ll}^3h_r^3 - 5h_{ll}^2h_r^4))/(h_{rr}(h_l + h_{rr})(h_{ll} + h_{rr})(h_r - h_{rr})(- 10h_l^4h_{ll}h_r + 5h_l^4h_{ll}h_{rr} - 15h_l^4h_r^2 + 10h_l^4h_rh_{rr} + 8h_l^3h_{ll}^2h_r - 4h_l^3h_{ll}^2h_{rr} - 13h_l^3h_{ll}h_r^2 + h_l^3h_{ll}h_rh_{rr} + 4h_l^3h_{ll}h_{rr}^2 - 33h_l^3h_r^3 + 13h_l^3h_r^2h_{rr} + 8h_l^3h_rh_{rr}^2 + 19h_l^2h_{ll}^2h_r^2 - 7h_l^2h_{ll}^2h_rh_{rr} - 3h_l^2h_{ll}^2h_{rr}^2 + 13h_l^2h_{ll}h_r^3 - 20h_l^2h_{ll}h_r^2h_{rr} + 7h_l^2h_{ll}h_rh_{rr}^2 - 15h_l^2h_r^4 - 13h_l^2h_r^3h_{rr} + 19h_l^2h_r^2h_{rr}^2 + 8h_lh_{ll}^2h_r^3 + 7h_lh_{ll}^2h_r^2h_{rr} - 9h_lh_{ll}^2h_rh_{rr}^2 + 10h_lh_{ll}h_r^4 + h_lh_{ll}h_r^3h_{rr} - 7h_lh_{ll}h_r^2h_{rr}^2 - 10h_lh_r^4h_{rr} + 8h_lh_r^3h_{rr}^2 + 4h_{ll}^2h_r^3h_{rr} - 3h_{ll}^2h_r^2h_{rr}^2 + 5h_{ll}h_r^4h_{rr} - 4h_{ll}h_r^3h_{rr}^2)) $,
\\
\\
and, $s^2_0 = -\left(s^2_{-2} + s^2_{-1}  + s^2_{1} + s^2_{2} \right).$
\\
\\
\noindent The coefficients of sixth order non-uniform grid based compact scheme for fourth derivative is obtained from Eq. \eqref{der4}.
The Taylor series expansion is used in Eq. \eqref{der4} and coefficients of various order are matched to obtain a linear relation between $p^4_{-1}$, $p^4_{1}$, $s^4_{-2}$, $s^4_{-1}$, $s^4_{0}$, $s^4_{1}$ and $s^4_{2}$, which is given as

\begin{equation}
    s^4_{-2} + s^4_{-1} + s^4_0 + s^4_{1} + s^4_{2} = 0,
\end{equation}
and

\begin{equation}
\left[\begin{array}{cccccc} -h_{\mathrm{ll}} & -h_{l} & h_{r} & h_{\mathrm{rr}} & 0 & 0\\ \frac{{h_{\mathrm{ll}}}^2}{2} & \frac{{h_{l}}^2}{2} & \frac{{h_{r}}^2}{2} & \frac{{h_{\mathrm{rr}}}^2}{2} & 0 & 0\\ -\frac{{h_{\mathrm{ll}}}^3}{6} & -\frac{{h_{l}}^3}{6} & \frac{{h_{r}}^3}{6} & \frac{{h_{\mathrm{rr}}}^3}{6} & 0 & 0\\ \frac{{h_{\mathrm{ll}}}^4}{24} & \frac{{h_{l}}^4}{24} & \frac{{h_{r}}^4}{24} & \frac{{h_{\mathrm{rr}}}^4}{24} & -1 & -1\\ -\frac{{h_{\mathrm{ll}}}^5}{120} & -\frac{{h_{l}}^5}{120} & \frac{{h_{r}}^5}{120} & \frac{{h_{\mathrm{rr}}}^5}{120} & h_{l} & -h_{r}\\ \frac{{h_{\mathrm{ll}}}^6}{720} & \frac{{h_{l}}^6}{720} & \frac{{h_{r}}^6}{720} & \frac{{h_{\mathrm{rr}}}^6}{720} & -\frac{{h_{l}}^2}{2} & -\frac{{h_{r}}^2}{2} \end{array}\right]
\left[\begin{array}{c} s^4_{-2}\\ s^4_{-1}\\ s^4_{1}\\ s^4_{2}\\ p^4_{-1}\\ p^4_{1} \end{array}\right] =
\left[\begin{array}{c} 0\\ 0\\ 0\\ 1\\ 0\\ 0 \end{array}\right].
\end{equation}

\noindent The solution of this linear system of equations is provided as,
\\
\\
\noindent $ p^4_{-1} = (h_r(h_l^2 + h_lh_{ll} + 2h_lh_r - h_lh_{rr} + h_{ll}^2 + 2h_{ll}h_r - h_{ll}h_{rr} - 2h_r^2 - 2h_rh_{rr} + h_{rr}^2))/((h_l + h_r)(2h_l^2 + 2h_lh_{ll} + 10h_lh_r - 2h_lh_{rr} - h_{ll}^2 - 2h_{ll}h_r + h_{ll}h_{rr} + 2h_r^2 + 2h_rh_{rr} - h_{rr}^2))$,
\\
\\
\noindent $ p^4_{1} = (h_l(- 2h_l^2 - 2h_lh_{ll} + 2h_lh_r + 2h_lh_{rr} + h_{ll}^2 - h_{ll}h_r - h_{ll}h_{rr} + h_r^2 + h_rh_{rr} + h_{rr}^2))/((h_l + h_r)(2h_l^2 + 2h_lh_{ll} + 10h_lh_r - 2h_lh_{rr} - h_{ll}^2 - 2h_{ll}h_r + h_{ll}h_{rr} + 2h_r^2 + 2h_rh_{rr} - h_{rr}^2)) $,
\\
\\
\noindent $ s^4_{-2} = -(360h_lh_r)/(h_{ll}(h_{ll} + h_r)(h_{ll} + h_{rr})(h_l - h_{ll})(2h_l^2 + 2h_lh_{ll} + 10h_lh_r - 2h_lh_{rr} - h_{ll}^2 - 2h_{ll}h_r + h_{ll}h_{rr} + 2h_r^2 + 2h_rh_{rr} - h_{rr}^2))$,
\\
\\
\noindent $ s^4_{-1} = (360h_r)/((h_l + h_r)(h_l + h_{rr})(h_l - h_{ll})(2h_l^2 + 2h_lh_{ll} + 10h_lh_r - 2h_lh_{rr} - h_{ll}^2 - 2h_{ll}h_r + h_{ll}h_{rr} + 2h_r^2 + 2h_rh_{rr} - h_{rr}^2))$,
\\
\\
\noindent $ s^4_{1} = (360h_l)/((h_l + h_r)(h_{ll} + h_r)(h_r - h_{rr})(2h_l^2 + 2h_lh_{ll} + 10h_lh_r - 2h_lh_{rr} - h_{ll}^2 - 2h_{ll}h_r + h_{ll}h_{rr} + 2h_r^2 + 2h_rh_{rr} - h_{rr}^2)) $,
\\
\\
\noindent $ s^4_{2} =  -(360h_lh_r)/(h_{rr}(h_l + h_{rr})(h_{ll} + h_{rr})(h_r - h_{rr})(2h_l^2 + 2h_lh_{ll} + 10h_lh_r - 2h_lh_{rr} - h_{ll}^2 - 2h_{ll}h_r + h_{ll}h_{rr} + 2h_r^2 + 2h_rh_{rr} - h_{rr}^2)) $,
\\
\\
and, $s^4_0 = -\left(s^4_{-2} + s^4_{-1}  + s^4_{1} + s^4_{2} \right).$

The coefficients of sixth order non-uniform grid based compact scheme for first derivative is already given in Ref. \cite{B96} and not reproduced here. The third derivative is obtained by taking the second derivative of the first derivative, obtained using non-uniform compact schemes following the expressions given after Eq. \eqref{der2}.

\end{document}